\documentclass[a4paper,12pt]{article}
\usepackage{graphicx,rotating,hyperref,slashed,amssymb,cancel,cite}
\pdfoutput=1
\usepackage{charter}

\newcommand{\bp}{\bar M_{\rm Pl}}
\hypersetup{colorlinks,bookmarksopen,bookmarksnumbered,
linkcolor=blus,pdfstartview=FitH,urlcolor=rossos,citecolor=verde}

\newcommand{\dxi}{\zeta}

\newcommand{\fig}[1]{~\ref{fig:#1}}
\newcommand{\gs}{f_0}
\newcommand{\gt}{f_2}
\newcommand{\Z}{\mathbb{Z}}
\usepackage{multicol}
\usepackage{color}
\definecolor{rosso}{cmyk}{0,1,1,0.4}
\definecolor{rossos}{cmyk}{0,1,1,0.55}
\definecolor{rossoc}{cmyk}{0,1,1,0.2}
\definecolor{blu}{cmyk}{1,1,0,0.3}
\definecolor{blus}{cmyk}{1,1,0,0.6}
\definecolor{bluc}{cmyk}{1,1,0,0.1}
\definecolor{verde}{cmyk}{0.92,0,0.59,0.25}
\definecolor{verdec}{cmyk}{0.92,0,0.59,0.15}
\definecolor{verdes}{cmyk}{0.92,0,0.59,0.4}

\newcommand{\sfrac}[2]{#1/#2}

\makeatletter

\def\hhref#1{\href{http://arxiv.org/abs/#1}{arXiv:#1}}
\usepackage{xstring} 
\newcommand{\hhrefq}[1]{\IfSubStr{#1}{:}{\href{http://inspirehep.net/search?ln=en&ln=en&p=#1&of=hb&action_search=Search&sf=&so=d&rm=&rg=25&sc=0}{InSpires:#1}}{\hhref{#1}}}

\def\art{\@ifnextchar[{\eart}{\oart}}
\def\eart[#1]#2#3#4#5#6{{\rm #2}, {\em #3 \bf #4} {\rm (#6) #5} ({\em #1})}
\def\article{\@ifnextchar[{\earticle}{\oarticle}}
\def\oarticle#1#2#3#4#5#6{{\rm #1}, {\sl ``#6''}, {\rm #2 #3 (#5) #4}}
\def\earticle[#1]#2#3#4#5#6#7{{\rm #2}, {\sl ``#7''}, {\rm #3 #4 (#6) #5}  [\hhrefq{#1}]}
\def\hepart[#1]#2{{\rm #2, \sl#1}}
\def\heparticle[#1]#2#3{#2, {\sl ``#3''} [\hhrefq{#1}]}
\newcommand{\doi}[1]{\href{http://dx.doi.org/#1}{[link]}}

\newcommand{\riga}[1]{\noalign{\hbox{\parbox{\textwidth}{#1}}}\nonumber}

\newcommand{\mub}{\bar{\mu}}
\renewcommand\&{&}

\newcommand{\eq}[1]{~{\rm (\ref{eq:#1})}}

\newcommand{\GeV}{\,{\rm GeV}}

\newcommand{\Tr}{\,{\rm Tr}}

\def\circa#1{\,\raise.3ex\hbox{$#1$\kern-.75em\lower1ex\hbox{$\sim$}}\,}

\newcommand{\beq}{\begin{equation}}
\newcommand{\eeq}{\end{equation}}

\newcommand{\bea}{\begin{eqnarray}}
\newcommand{\eea}{\end{eqnarray}}
\newcommand{\be}{\begin{equation}}
\newcommand{\ee}{\end{equation}}
\def\ba{\begin{array} }
\def\bac{\begin{array} {c}}
\def\bacc{\begin{array} {cc}}
\def\baccc{\begin{array} {ccc}}
\def\bacccc{\begin{array} {cccc}}
\def\ea{\end{array}}
\def\bea{\begin{eqnarray}}
\def\eea{\end{eqnarray}}
\font\tenrsfs=rsfs10 at 12pt
\font\sevenrsfs=rsfs7
\font\fiversfs=rsfs5
\newfam\rsfsfam
\textfont\rsfsfam=\tenrsfs
\scriptfont\rsfsfam=\sevenrsfs
\scriptscriptfont\rsfsfam=\fiversfs
\def\mathscr#1{{\fam\rsfsfam\relax#1}}

\def\Lag{\mathscr{L}}

\def\circa#1{\,\raise.3ex\hbox{$#1$\kern-.75em\lower1ex\hbox{$\sim$}}\,}

\def\hhref#1{\href{http://arxiv.org/abs/#1}{arXiv:#1}} 

\newcounter{alphaequation}[equation]
\def\thealphaequation{\theequation\hbox to
0.6em{\hfil\alph{alphaequation}\hfil}}
\def\eqnsystem#1{
\def\@eqnnum{{\rm (\thealphaequation)}}
\def\@@eqncr{\let\@tempa\relax \ifcase\@eqcnt \def\@tempa{& & &} \or
  \def\@tempa{& &}\or \def\@tempa{&}\fi\@tempa
  \if@eqnsw\@eqnnum\refstepcounter{alphaequation}\fi
\global\@eqnswtrue\global\@eqcnt=0\cr}
\refstepcounter{equation} \let\@currentlabel\theequation \def\@tempb{#1}
\ifx\@tempb\empty\else\label{#1}\fi
\refstepcounter{alphaequation}
\let\@currentlabel\thealphaequation
\global\@eqnswtrue\global\@eqcnt=0 \tabskip\@centering\let\\=\@eqncr
$$\halign to \displaywidth\bgroup \@eqnsel\hskip\@centering
$\displaystyle\tabskip\z@{##}$&\global\@eqcnt\@ne
\hskip2\arraycolsep\hfil${##}$\hfil& \global\@eqcnt\tw@\hskip2\arraycolsep
$\displaystyle\tabskip\z@{##}$\hfil
\tabskip\@centering&\llap{##}\tabskip\z@\cr}
\def\endeqnsystem{\@@eqncr\egroup$$\global\@ignoretrue} \makeatother

\newcommand{\SU}{\,{\rm SU}}

\begin{document}

\vspace{1cm}

\begin{center}
CERN-TH-2017-105\hfill IFUP-TH/2017

\vspace{0.7cm}
{\LARGE \bf \color{rossos}
Agravity  up to infinite energy  }\\[1cm]

{\large\bf Alberto Salvio$^{a}$ {\rm and} Alessandro Strumia$^{a,b}$}  
\\[7mm]
{\it $^a$ } {\em CERN, Theoretical Physics Department, Geneva, Switzerland}\\[1mm]
{\it $^b$ Dipartimento di Fisica dell'Universit{\`a} di Pisa and INFN, Italy}

\vspace{1cm}
{\large\bf\color{blus} Abstract}
\begin{quote}
The self-interactions of the  conformal mode of the graviton are controlled, in dimensionless gravity  theories (agravity),
by a  coupling $f_0$ that is not asymptotically free.
We show that, nevertheless, agravity can be a complete theory valid up to infinite energy.
When $f_0$ grows to large values, the conformal mode of the graviton decouples from the rest of the theory
and does not hit any Landau pole provided that scalars are asymptotically conformally coupled and all other couplings approach fixed points.
Then, agravity can flow to conformal gravity at infinite energy. 
We identify scenarios where the Higgs mass does not receive unnaturally large physical corrections.
We also show a useful equivalence between agravity and conformal gravity plus two extra conformally coupled scalars, and give
a  simpler form for the renormalization group equations of dimensionless couplings as well as of massive parameters 
in the presence of the most general matter sector.

\end{quote}

\thispagestyle{empty}


{\parbox{0.89\textwidth}{{\small \tableofcontents}}}
\end{center}

\setcounter{footnote}{0}

\section{Introduction}

The idea that scalars, like the Higgs, must be accompanied by new physics that protects their lightness
from power-divergent quantum corrections
led to the following view of mass scales in nature: 
the weak scale is the supersymmetric scale, and the Planck scale is the string scale.
The non-observation of supersymmetric particles around the weak scale
challenged this  scenario, leading to the alternative idea
that only physical corrections to scalar masses must satisfy naturalness.
Namely, extra new particles with mass $M_{\rm extra}$ and coupling $g_{\rm extra}$ to the Higgs, must satisfy
\beq \delta M_h \sim g_{\rm extra} M_{\rm extra} \circa{<} M_h.\eeq
A rationale for ignoring power-divergent corrections is the following.
The one-loop quantum correction to the masses of
scalars, vectors and of the graviton is power divergent, showing
the dangers of attributing physical meaning to power-divergent corrections.
A cut-off (such as string theory) that knows  that vector and graviton masses are
protected by gauge invariance can keep them to zero, while giving a large correction to scalar masses.
A less smart cut-off (such as dimensional regularization) can be blind to the difference,
and set to zero all power divergences.
The simplest cut-off with this property is no cut-off: 
a  theory where all renormalizable couplings
flow up to infinite energy without hitting Landau poles.

The above arguments motivate the following scenario:
if nature is described at fundamental level by a dimensionless Lagrangian, 
all power-divergent quantum corrections --- being dimensionful --- must be interpreted as vanishing.
Taking gravity into account, the most general dimensionless action in $3+1$ space-time dimensions
contains gauge couplings, Yukawa couplings, scalar quartics,
{non-minimal $\xi$-couplings between scalars and gravity}
 and, in the purely gravitational sector, two dimensionless gravitational couplings, $f_0$ and  $f_2$, analogous to gauge couplings:
\beq \label{eq:Ladim}
S=\int d^4x\,\sqrt{|\det g|} \bigg[ \frac{R^2}{6\gs^2} + \frac{\frac13 R^2 -  R_{\mu\nu}^2}{\gt^2}  + \Lag_{\rm matter}\bigg],
\eeq
where $\Lag_{\rm matter}$ corresponds to the part of the Lagrangian that depends on the matter fields, with dimensionless parameters only. This theory~\cite{Utiyama:1962sn} is renormalizable, as suggested in~\cite{Weinberg:1974tw} and  formally proven in~\cite{Stelle:1976gc}.
The weak scale, the QCD scale and the Planck scale can be dynamically generated~\cite{agravity}
from vacuum expectation values or from condensates.
Perturbative dimensionless theories automatically give slow-roll inflation~\cite{agravity,WetterichInf,1502.01334,Farzinnia:2015fka,Salvio:2017xul} (see also refs.~\cite{Salles:2014rua,Ivanov:2016hcm} for related studies).

However, eq.\eq{Ladim} means that 4 derivatives act on the graviton: 
thereby some graviton components have a negative kinetic term.\footnote{This can maybe be
avoided introducing an infinite series of higher derivative terms~\cite{Biswas:2005qr}, but the resulting gravity theories 
contain infinite free parameters and are not known to be renormalizable.}
Classically the theory in (\ref{eq:Ladim})  is sick~\cite{Ostro}:  the energy is unbounded from below.
A sensible quantum theory might exist, analogously
to what happens with fermions: their classical energy is negative, but their quantum theory is sensible.\footnote{The ample literature of `ghosts'
was critically reviewed in \cite{1512.01237}, where it was proposed that 
a 4-derivative variable $q(t)$ contains two canonical degrees of freedom (d.o.f.),
$q_1 = q$ and $q_2 = \dot q$, with opposite time-reflection parity,
such that usual $T$-even representation
({$q_1 |x\rangle = x  |x\rangle$ and $p_1|x\rangle =i\frac{d}{dx}|x\rangle$}
) must be combined with the $T$-odd
representation 
({$q_2 |y\rangle = i y  |y\rangle$ and $p_2|y\rangle =\frac{d}{dy}|y\rangle$}
)
obtaining consistent results
(positive energy, normalisable wave functions, Euclidean continuation),
although the interpretation of the resulting negative norm is unclear.}
We will not address this problem here.

\medskip

We will here study whether  this theory can flow up to infinite energy.
The Quantum Field Theory (QFT) part can have this property.
Realistic TeV-scale extensions of the Standard Model (SM) can be asymptotically free~\cite{trini},
and it is not known whether the SM itself can be asymptotically safe, in a non-perturbative regime~\cite{1701.01453}.
The gravitational coupling $f_2$ is asymptotically free.
The difficulty resides in the coupling $f_0$: a small $f_0$ grows with energy, until it becomes large.

In this paper we will show that, despite this, the theory can flow up to infinite energy, in an unusual way.
In section~\ref{agravtysec} we  present an alternative formulation of agravity that makes it easier to compute its renormalization group equations (RGE):
$f_0$ becomes the quartic of a special scalar, the conformal mode of the agraviton. 
Then, a large $f_0$ means that the conformal mode of the agraviton  gets strongly self-coupled.
The rest of the theory decouples from it, if at the same time all scalars become conformally coupled, 
namely if all $\xi$ parameters run to $-1/6$, {and all the other couplings reach ultraviolet (UV) fixed points}, where all $\beta$-functions vanish.

In section~\ref{sigma} we isolate the conformal mode of the graviton and show that 
its strong dynamics is such that $f_0$ does not hit  a Landau pole.
This means that the infinite-energy limit of agravity can be conformal gravity.
The unusual phenomenon that allows to reach infinite energy is that the conformal mode of the graviton
fluctuates freely, but the rest of theory is not coupled to it:
it becomes a gauge redundancy of a new local symmetry, Weyl symmetry.
Since this symmetry is anomalous, conformal gravity cannot be the {complete}
theory: 
going to lower energy the conformal model of the graviton starts coupling to the rest of the theory, which becomes agravity.
This issue is discussed in  section~\ref{conformal}.
In section~\ref{nat} we propose scenarios where the Higgs mass does not receive unnaturally large corrections.
Conclusions are given in section~\ref{concl}. Finally, in the appendix we provide a new and simple expression for the one-loop RGE of all dimensionless parameters (appendix \ref{RGE}) as well as of all dimensionful parameters (appendix \ref{RGEM}) in the presence of the most general matter sector, which was not studied before.

%

\section{Agravity}\label{agravtysec}
Allowing for generic scalars $\phi_a$ with generic dimensionless coupling $\xi_{ab}$ to gravity, $-\frac12 \xi_{ab}\phi_a\phi_b R$,
the one-loop RGE for $f_0$ is~\cite{Avramidi:1985ki,Avramidi:1986mj,deBerredoPeixoto:2004if,agravity} 
\beq
(4\pi)^2\frac{d \gs^2 }{d\ln\mub}= \frac53 \gt^4 + 5 \gt^2 \gs^2 + \frac56 \gs^4 +\frac{\gs^4}{12} (\delta_{ab}+6\xi_{ab})(\delta_{ab}+6\xi_{ab}) >0 
\qquad \hbox{for $f_0 \ll 1$,}
\label{eq:RGEf0}
\eeq
where $\mub$ is the renormalization scale in the modified minimal subtraction scheme (see also \cite{Julve:1978xn,Frad} for a previous attempt to determine this RGE).
This shows that, in all theories, $f_0$ is asymptotically free {only} for $f_0^2<0$.
However, negative $f_0^2$ corresponds to a run-away potential~\cite{agravity,1502.01334},
and this instability cannot be made harmless (or even beneficial for explaining dark energy)
by invoking a small enough negative $f_0^2$, since tests of gravity exclude extra graviton components below 0.05 eV (see~\cite{Einhorn1} for attempts to have $f_0^2 < 0$). The fact that $f_0^2<0$ is phenomenologically problematic was already noted in~\cite{agravity}, where it was pointed out that it leads to a tachyonic instability.
Barring 
 stabilisation through background effects in cosmology, one needs $f_0^2>0$.
But the one-loop RGE show that
a small  $f^2_0>0$ grows until it becomes non-perturbative.\footnote{Different statements in the literature (even recent)
appear either because  some previous results contained wrong signs
or because some authors use computational techniques that try to give a physical meaning to power divergences,
obtaining gauge-dependent and cut-off dependent results.
Claims that a run-away potential with very small $f_0$ can mimic Dark Energy do not take into account bounds on extra graviton components.}

\begin{table}[t]
$$\begin{array}{cc|ccccc}
&& \hbox{dilatation} &\otimes &\hbox{diffeomorphism} &=& \hbox{Weyl transformation}\\ \hline
\hbox{coordinates} & dx^\mu  &  e^\sigma dx^\mu &&e^{-\sigma} dx^\mu && dx^\mu\\
\hbox{graviton} & g_{\mu\nu} &g_{\mu\nu} && e^{2\sigma} g_{\mu\nu} &&e^{2\sigma}  g_{\mu\nu} \\
\hbox{scalars} &\phi & e^{-\sigma} \phi && \phi &&e^{-\sigma} \phi \\
\hbox{vectors} &  V_{\mu}
 & e^{-\sigma}  V_{\mu}
 && e^{\sigma}  V_{\mu}
&&V_{\mu}
\\
\hbox{fermions} & \psi & e^{-3\sigma/2} \psi &&\psi &&  e^{-3\sigma/2} \psi
\end{array}
$$
\caption{\em Transformations of coordinates and fields under a Weyl transformation.\label{tab:Weyl}}
\end{table}%

These RGE show peculiar features.
Only scalars (not vectors nor fermions) generate $f_0$ at one-loop, and
only if their $\xi$-couplings have a non-conformal value, $\xi_{ab}\neq -\delta_{ab}/6$.
The $\xi$-couplings often appear in the RGE in the combination $\xi_{ab}+\delta_{ab}/6$, but not always.
The coupling $f_0$ appears at the denominator in the RGE for the $\xi$-couplings~\cite{agravity}.

The above features can be understood noticing that a new symmetry appears in the limit $f_0 \rightarrow \infty$ and $\xi_{ab} \rightarrow -\delta_{ab}/6$:
the Weyl (or   local  conformal) symmetry.
The Weyl symmetry is a {local} dilatation $dx^\mu\to e^{\sigma(x)} dx^\mu$ compensated by the
special diffeomorphism $dx^\mu \to e^{-\sigma(x)} dx^\mu$ such that the coordinates $dx^\mu$ remain unaffected.
The various fields rescale under a dilatation as determined by their mass dimension,
and transform under a diffeomorphism as dictated by their Lorentz indices, as summarized in table~\ref{tab:Weyl}.

Agravity is invariant under global Weyl transformations:
being dimensionless, it is invariant under global dilatations (for which $\sigma$ does not depend on $x$);
being covariant, it is invariant under local diffeomorphisms.

Agravity is not invariant under local Weyl transformations.
A generic dimensionless theory can be written in terms of the metric $g_{\mu\nu}$, real scalars $\phi_a$, Weyl fermions $\psi_j$ and vectors $V_A$ (with field strength $F_{\mu\nu}^A$).
The action $S = \int d^4x \sqrt{|\det g|}\, \Lag$ can be split as  $S = S_{\rm Weyl} + S_{\cancel{\rm Weyl}}$
where the first part is invariant under Weyl transformations
\bea \label{eq:Lconf}
\Lag_{\rm Weyl} &=&  \frac{\frac13 R^2 -  R_{\mu\nu}^2}{\gt^2} 
- \frac14 (F_{\mu\nu}^A)^2 + \frac{(D_\mu \phi_a)^2}{2}  + \bar\psi_j i\slashed{D} \psi_j  +\nonumber\\
&& + \frac{1}{12} \phi_a^2 R - \frac12 (Y^a_{ij} \psi_i\psi_j \phi_a + \hbox{h.c.}) - \frac{\lambda_{abcd}}{4!} \phi_a\phi_b\phi_c\phi_d
\label{eq:Lgen}\eea
and the second part
\beq  \label{eq:Lnotconf}
\Lag_{\cancel{\rm Weyl}} =  \frac{R^2}{6f_0^2}  -  \frac12 \dxi_{ab} \phi_a \phi_b R,\qquad \hbox{where } \quad\zeta_{ab}\equiv\xi_{ab} +\delta_{ab}/6\eeq
is not invariant.\footnote{We omitted  the topological Gauss-Bonnet  term.}  
To see this we will now perform a Weyl transformation 
\be g_{\mu\nu}(x) \to  e^{2\sigma(x)} g_{\mu\nu}(x),\quad \phi(x)\rightarrow e^{-\sigma(x)}\phi(x), \quad \psi(x)\rightarrow e^{-3\sigma(x)/2}\psi(x), \qquad V_\mu\rightarrow V_\mu. \label{sigmaIntro}\ee
 This will also lead to an equivalent formulation of the theory.

\subsubsection*{Equivalent formulations of agravity}

The extra scalar field $\sigma(x)$, defined in (\ref{sigmaIntro}), will be called the `conformal mode of the agraviton';
for the moment it is introduced as an extra gauge redundancy. {We will comment on the corresponding gauge symmetry later on.}



{All terms in eq.\eq{Lconf} are invariant under Weyl transformations. Since  vectors and fermions  appear only in eq.\eq{Lconf}, $\sigma$ does not couple to them.
Only the terms  that break Weyl symmetry give rise to interactions of $\sigma$.}
Transformation (\ref{sigmaIntro})  leads to 
\beq \label{eq:dR}
\sqrt{|\det g|} \rightarrow e^{4\sigma}\sqrt{|\det g|},\qquad
R \rightarrow e^{-2\sigma} (R - 6 e^{-\sigma}\Box e^\sigma).\eeq
Therefore, the Weyl-breaking part of the Lagrangian becomes
\beq \label{eq:Lnotconfsigma}
\sqrt{|\det g|} \Lag_{\cancel{\rm Weyl}} = \sqrt{|\det g|} \left[  \frac{(R - 6 e^{-\sigma}\Box e^{\sigma})^2}{6f_0^2}  -  \frac12 \dxi_{ab} \phi_a \phi_b (R - 6 e^{-\sigma}\Box e^{\sigma})\right], \label{eq1}\eeq
{which is one simple way to rewrite $\Lag_{\cancel{\rm Weyl}}$ that will be used later on. 

Another simple and useful form of $\Lag_{\cancel{\rm Weyl}}$ can be obtained from (\ref{eq1}) as follows.}
We define $\Omega_L =e^\sigma$ and complete the square rewriting eq.\eq{Lnotconfsigma} as
\beq \Lag_{\cancel{\rm Weyl}} =  \frac{A^2}{6f_0^2}  -  \frac38 f_0^2 (\dxi_{ab} \phi_a \phi_b)^2,\qquad
A=R - 6 \frac{\Box \Omega_L}{\Omega_L}  -\frac32  f_0^2 \dxi_{ab} \phi_a\phi_b.\eeq
Next we write the square as $A^2/6f_0^2 = -\frac16 f_0^2 \Omega_L^2\Omega_R^2 +\frac13 \Omega_R \Omega_L A$
by introducing an auxiliary field $\Omega_R$ with quadratic action, such that integrating it out gives back the original action.
The resulting expression only contains the combination
$\Omega_L\Omega_R$ that is invariant under 
$\Omega_L\to t\Omega_L$, $\Omega_R\to\Omega_R/t$,
which forms a SO(1,1) scale symmetry.
Indeed, one can verify that SO(1,1) is broken by adding Lagrangian terms with dimensionful coefficients,
such as the Einstein-Hilbert term or the cosmological constant, as done later in eq.\eq{notSO11}.
Now, we can rewrite $\Omega_L \Omega_R$ in vectorial notation as
$\Omega_L\Omega_R = \frac1{4} (\Omega_+^2 - \Omega_-^2) =\frac1{4} \vec \Omega^2$ 
by  going from the  ``light-cone basis'' $\Omega_{L,R}$ to the $\Omega_\pm$ basis as
$\Omega_L = t( \Omega_+  - \Omega_-)/{2}$ and $\Omega_R=(\Omega_+ +  \Omega_-)/{2t}$
and defining the SO(1,1) vector  $\vec \Omega = (\Omega_+, \Omega_-)$.
Then the Weyl-breaking part of the action can be written in the final form
\beq \label{eq:Snotconf2}
S_{\cancel{\rm Weyl}} =  \int d^4x\sqrt{|\det g|}\bigg[ \frac{g^{\mu\nu}}{2} (\partial_\mu\vec\Omega)(\partial_\nu\vec\Omega)+
\frac{1}{12} \vec\Omega^2 R
-\frac{f_0^2}{96}\bigg(\vec\Omega^2 +6\zeta_{ab}\phi_a\phi_b\bigg)^2\bigg]. \eeq
The non-trivial result is that {\em the Weyl-breaking part of the action has been rewritten as an extra Weyl-invariant action} involving the
extra scalar  SO(1,1) doublet $\vec \Omega$, which describes the conformal mode of the agraviton. 

We have not (yet)  imposed any constraint on the metric $g_{\mu\nu}$ after the transformation in eq.~(\ref{sigmaIntro}); 
therefore we have a Weyl-type gauge invariance acting as 
\beq g_{\mu \nu}(x) \rightarrow e^{-2\chi(x)}g_{\mu\nu}(x), \quad
\phi(x)\rightarrow e^{\chi(x)}\phi(x), \quad \psi(x)\rightarrow e^{3\chi(x)/2}\psi(x), \qquad V_\mu\rightarrow V_\mu \eeq
where $\chi(x)$ is an arbitrary real function of $x$.
The transformation $\sigma \to \sigma +\chi$ is equivalent to including $\Omega_L =e^\sigma$ and $\Omega_R$ among the scalars $\phi$.
Therefore, agravity is equivalent to conformal gravity plus two extra conformally coupled scalars, $\Omega_+$ and $\Omega_-$.\footnote{Similar remarks have been made in the context of Einstein gravity (rather than in agravity) in~\cite{Hooft:2010nc,tHooft:2011aa,Hooft:2015rdz}, where  it was found that Einstein gravity is equivalent to conformal gravity plus a single conformally coupled scalar. Other similar statements have been made in a different theory without the $R^2/6f_0^2$ term in~\cite{Hamada:2002cm,Hamada:2009hb}.}
In the new  formulation of agravity with the field $\vec\Omega$, the gravitational couplings $f_0$ and $\xi_{ab}$ have become scalar quartic couplings.

The formulations presented in this section certainly are equivalent at the classical level. 
At quantum level, the equivalence needs to take into account
the anomalous transformation law of the path-integral measure, which amounts
to adding an effective $\sigma$-dependent term in the action.
This amounts to say that  $\sigma$ starts coupling to terms that break scale invariance proportionally to their quantum $\beta$-functions.
These extra couplings only affect RGEs at higher loop orders, as we will discuss in section \ref{conformal}.

It is now clear why the one-loop RGE for $f_0$, eq.\eq{RGEf0},
 does not receive contributions from fermions and vectors: $f_0^2$
is the quartic coupling of a neutral scalar with no Yukawa interactions.
A positive $f_0^2$ corresponds to a positive quartic.  
Furthermore the symmetry SO(1,1) can be complexified into SO(2) by redefining $\Omega_- \to i \Omega_-$
without affecting the RGE at perturbative level: only non-perturbative large field fluctuations are sensitive to the difference.
By defining an extended set of quartic couplings, $\lambda_{ABCD}$, 
where the capital indices
 run such that the quartics that involve the two extra scalars $\vec\Omega$ are included,
the generic RGE for the scalar quartics only, known in a generic QFT up to two loops, are~\cite{Luo:2002ti}
\beq \label{eq:tildelambdaRGE}
\frac{d \lambda_{ABCD}}{d\ln\bar\mu} = \frac{1}{(4\pi)^2} \sum_{\rm perms}\frac18 \lambda_{ABEF}\lambda_{EFCD}+
\frac{1}{(4\pi)^4}\bigg[\frac{\gamma}{2} \lambda_{ABCD}- \frac{1}{4}\sum_{\rm perms} \lambda_{ABEF}\lambda_{CEGH}\lambda_{DFGH}
 \bigg]+\cdots,\eeq
where $\gamma = \Lambda_{AA}+\Lambda_{BB}+\Lambda_{CC} + \Lambda_{DD} $
(with $\Lambda_{AB} = \frac16 \lambda_{ACDE}\lambda_{BCDE}$)
is the scalar wave-function renormalization,
 the sums  run over the $4!$ permutations of $ABCD$  and $\cdots$ is the contribution of the other couplings.

From eq.\eq{tildelambdaRGE} one can re-derive the one-loop RGE for $f_0$ and $\xi_{ab}$, computed as gravitational couplings in~\cite{agravity}.
The two results agree.
Furthermore, the same RGE acquire a simpler form if rewritten in terms of the $\lambda_{ABCD}$ coefficients.
The RGE are explicitly written in eq.~(\ref{sys:agravityRGE2}) in appendix~\ref{RGE}, 
and neither $f_0$ nor any other coupling appears anymore at the denominator in the RGE.

\medskip

\subsubsection*{The graviton propagator}
A gravitational computation is now only needed to compute the part of the RGE involving $f_2$.
So far the field $\sigma$, or $\vec \Omega$, has been introduced as an extra gauge redundancy.
One can fix it by setting $\sigma=0$, going back to the original formulation where the full RGE were computed in~\cite{agravity}. 
In the rest of this section (which contains technical details used only for a double check of the main results)
we show how one can choose an alternative convenient condition: 
that the fluctuation $h'_{\mu\nu}$ around the flat space of $g_{\mu\nu}$ {\it after} the transformation in eq.~(\ref{sigmaIntro}) has vanishing trace, that is 
  \beq  h' \equiv  \eta^{\mu\nu}h'_{\mu \nu} = 0. \label{DefSigma} \eeq
  We have introduced a prime in $h'_{\mu \nu}$ to distinguish it from the fluctuation $h_{\mu\nu}$ around the flat space of the metric {\it before} transformation (\ref{sigmaIntro}). 
  The new variables $h'_{\mu\nu}$ and $\sigma$ are given in terms of the old  ones (the trace $h\equiv \eta^{\mu\nu} h_{\mu\nu}$ and the traceless part $h^{\rm TL}_{\mu\nu}\equiv h_{\mu\nu}-\eta_{\mu\nu}h/4$) by
\beq  e^{2\sigma} = 1+\frac{h}{4}, \qquad h'_{\mu\nu} = e^{-2\sigma}h^{\rm TL}_{\mu\nu}. \label{FieldRedef2}
\eeq
The path integral measure $Dg_{\mu\nu}\equiv Dh \, Dh^{\rm TL}_{\mu\nu}$  splits as
$Dg_{\mu\nu} = Dh'_{\mu\nu}\,D\sigma = Dh'_{\mu\nu} D\vec\Omega$.
We  neglect here the Weyl anomaly because, as explained above,  it does not affect the one-loop RGE. 


\medskip

In order to compute quantum effects, we consider the following convenient gauge-fixing for the diffeomorphisms $x^\mu \to x^\mu+\xi^\mu(x)$:
\be \partial^{\mu}h'_{\mu\nu}=0, \label{eq:gaugeCond2}\ee
where we use the flat metric $\eta_{\mu\nu}$ to raise and lower the indices.
This choice avoids kinetic mixing between $\sigma$ and $h'_{\mu\nu}$ and leads to a simple
%
propagator of $h'_{\mu\nu}$ 
\be D'_{\mu\nu\,\rho\sigma} = -2f_2^2 \frac{ i}{k^4} P^{(2)}_{\mu\nu\rho\sigma}, \label{Prophp}\ee
where
\be P^{(2)}_{\mu\nu\rho\sigma} = \frac12 T_{\mu\rho}T_{\nu\sigma} + \frac12 T_{\mu\sigma} T_{\nu\rho}-\frac13 T_{\mu\nu}T_{\rho \sigma}, \qquad  T_{\mu\nu} = \eta_{\mu\nu} - k_\mu k_\nu/k^2. \ee
To determine the Lagrangian of the Fadeev-Popov ghosts we have to perform the variation of $\partial^{\mu}h'_{\mu\nu}$ with respect to diffeomorphisms, whose effect on $h_{\mu\nu}$ at the linear level in $\xi^\mu$ is 
\be h_{\mu\nu} \rightarrow  h_{\mu\nu}
- (\partial_\mu \xi_\nu + \partial_\nu \xi_\mu) - (h_{\alpha\mu}\partial_\nu +h_{\alpha\nu}\partial_\mu + (\partial_\alpha h_{\mu\nu}))\xi^\alpha. \label{diffeo2}
\ee
The effect of diffeomorphisms on $h'_{\mu\nu}$ and $\sigma$ can be computed by first splitting
eq.~(\ref{diffeo2}) in its traceless and trace parts,
\bea h &\rightarrow& h -2 \partial_\mu \xi^\mu - 2 h^{\rm TL}_{\alpha\mu}\partial^\mu \xi^\alpha -\frac12 h \partial_\mu \xi^\mu- \xi^\alpha \partial_\alpha h, \\ 
h^{\rm TL}_{\mu\nu} &\rightarrow& h^{\rm TL}_{\mu\nu} -\partial_\nu \xi_\mu-\partial_\mu \xi_\nu+\frac12\eta_{\mu\nu} \partial_\alpha\xi^\alpha - h^{\rm TL}_{\alpha\mu} \partial_\nu \xi^\alpha- h^{\rm TL}_{\alpha\nu} \partial_\mu \xi^\alpha -  \partial_\alpha h^{\rm TL}_{\mu\nu} \xi^\alpha \nonumber \\
&&+\frac12 \eta_{\mu\nu} h^{\rm TL}_{\alpha\beta} \partial^\beta \xi^\alpha -\frac14 h (\partial_\nu\xi_\mu+\partial_\mu\xi_\nu)+\frac18 \eta_{\mu\nu} h \partial_\alpha \xi^\alpha ,   \eea
 and next by using eq.~(\ref{FieldRedef2}) to express $h^{\rm TL}_{\mu\nu}$ and $h$ in terms of $h'_{\mu\nu}$ and $\sigma$.  The result is 
\bea e^{2\sigma} &\rightarrow& e^{2\sigma} \left(1- \frac12 \partial_{\mu}\xi^\mu- \frac12 h'_{\mu\alpha}\partial^{\mu}\xi^\alpha -2\xi_\alpha \partial^\alpha \sigma\right), \label{diff1}\\ 
h'_{\mu\nu} &\rightarrow& h'_{\mu\nu} -\partial_\nu \xi_\mu-\partial_\mu \xi_\nu+\frac12\eta_{\mu\nu} \partial_\alpha\xi^\alpha - h'_{\alpha\mu} \partial_\nu \xi^\alpha- h'_{\alpha\nu} \partial_\mu \xi^\alpha -  \partial_\alpha h'_{\mu\nu} \xi^\alpha +\frac12 h'_{\mu\nu} \partial_\alpha \xi^\alpha \nonumber \\
&&+\frac12 \eta_{\mu\nu} h'_{\alpha\beta} \partial^\beta \xi^\alpha+\frac12 h'_{\mu\nu}h'_{\alpha\beta} \partial^\beta \xi^\alpha. \label{eq:dff2} \eea
Notice that the transformation law of $h'_{\mu\nu}$ is independent of $\sigma$:
having used the gauge in eq.~(\ref{eq:gaugeCond2})
the Fadeev-Popov procedure does not generate any new coupling of $\sigma$ to the Fadeev-Popov ghosts.\footnote{We treated the
Weyl transformation as a change of variables in field space.
We could equivalently have seen  it as an extra gauge redundancy.
In this alternative  formalism, using the Fadeev-Popov  procedure to fix both diffeomorphisms and the Weyl symmetry,
the gauge fixing in eq.\eq{gaugeCond2} avoids mixed terms in the ghost system;
the ghosts for the Weyl gauge fixing are non-dynamical and integrating them out 
is equivalent to the modified diffeomorphism transformation law of the traceless graviton, eq.\eq{dff2}.}
In conclusion, we have shown how to implement the gauge where the graviton is traceless.


\section{Conformal gravity}\label{conformal}
We return to our physical issue: the coupling $f_0$ is not asymptotically free.
In section~\ref{sigma} we will argue that $f_0$  grows with energy, becoming non-perturbative at $f_0 \sim 4\pi$ and continuing to grow up to
$f_0\to \infty$ in the limit of infinite energy,
such that the $R^2/6f_0^2$ term disappears from the action.
In this section we show that this limit is well defined.  
It is precisely defined as agravity with parameters chosen such that all Weyl-breaking terms
  $\Lag_{\cancel{\rm Weyl}} $ in eq.\eq{Lnotconf} vanish:
\beq \label{eq:conformal}
f_0 =\infty,\qquad \xi_{ab}  =- \frac{\delta_{ab}}{6} .\eeq
The $R^2/6 f_0^2$ term provides the kinetic term for $\sigma$, the conformal mode of the agraviton.
Thereby $\sigma$ fluctuates wildly in the limit $f_0 \to \infty$.
Indeed, the agraviton propagator of~\cite{agravity} has a contribution proportional to $f_0^2$,
which diverges as $f_0\to\infty$.
Faddeev and Popov have shown how to deal with these situations: 
add an extra gauge-fixing for the extra  gauge redundancy appearing in conformal gravity,
local Weyl transformations.

\medskip

In general, conformal gravity is not a consistent quantum theory, because its Weyl gauge symmetry is anomalous.
In a simpler language, the dimensionless couplings run with energy as described by their RGE.\footnote{One might hope that all couplings stay at fixed points at all energies, but this possibility is excluded because one must recover a non-conformal behaviour at low energies for phenomenological reasons.}
The theory is no longer scale invariant, and the conformal mode of the graviton couples to all non-vanishing $\beta$-functions.
The Weyl-breaking terms of the agravity Lagrangian are generated back by quantum corrections.
The consistent quantum theory is agravity. For this reason our work differs from articles where conformal gravity is proposed as a complete theory of gravity~\cite{Mannheim:2009qi,Mannheim:2011ds}.

Nevertheless, conformal gravity can be the consistent infinite-energy limit of agravity provided that
all $\beta$-functions vanish at infinite energy:
the theory must be asymptotically free or asymptotically safe, in other words all couplings other than $f_0$ have to reach a UV fixed point
where all $\beta$-functions vanish, as we will see.

In this section we clarify these issues by computing the one-loop RGE of conformal gravity coupled to a generic 
{matter sector}, as in eq.\eq{Lconf}.
The RGE can be obtained without performing any  extra computation by using the perturbative equality obtained in the previous section:
agravity is equivalent to conformal gravity plus two extra scalars, $\vec\Omega$.
In the other direction, this means that conformal gravity has the same RGE as agravity {\em minus} two scalars.
Thereby the RGE for $f_2$ in conformal gravity {is obtained by substituting $N_s\rightarrow N_s-2$ in (\ref{RGEf2})  obtaining}
\beq\label{eq:RGEf2c}
(4\pi)^2\frac{df_2^2}{d\ln\mub}=
-f_2^4\bigg(\frac{199}{15} +\frac{N_V}{5}+\frac{N_f}{20}+\frac{N_s}{60}
\bigg)\qquad \hbox{{(for $f_0\to \infty$ and $\xi_{ab}\rightarrow -\frac16 \delta_{ab}$}).
}\eeq
This reproduces the result obtained in~\cite{ShapiroAnom} with a dedicated computation in the gauge of eq.~(\ref{DefSigma}),
where only the traceless part of the graviton propagates, see eq.~(\ref{Prophp}).
Then, the one-loop RGE for all other parameters can be obtained from the agravity RGE, listed in the appendix,
by dropping those for $f_0$ and 
$\xi_{ab}$, as well as the terms involving $f_0$ and $\xi_{ab}+\delta_{ab}/6$ from the remaining RGE.
The result is
\bea\label{eq:RGEY2}
(4\pi)^2 \frac{dY^a}{d\ln\mub} &=& \frac12(Y^{\dagger b}Y^b Y^a + Y^a Y^{\dagger b}Y^b)+ 2 Y^b Y^{\dagger a} Y^b + \nonumber\\
&&+ Y^b \Tr(Y^{\dagger b} Y^a) - 3 \{ C_{2F} , Y^a\}  + \frac{15}{8}\gt^2 Y^a,~~~ \\
(4\pi)^2 \frac{d\lambda_{abcd}}{d\ln\mub} &=&  \sum_{\rm perms} \bigg[\frac18
  \lambda_{abef}\lambda_{efcd}+
  \frac38 \{\theta^A,\theta^B\}_{ab}\{\theta^A ,\theta^B\}_{cd}
  -\Tr\, Y^a Y^{\dagger b} Y^c Y^{\dagger d}+
   \nonumber
\\
&&\label{eq:RGElambda2}
+\frac5{288} \gt^4 \delta_{ab}\delta_{cd}+ \lambda_{abcd} \bigg[ \sum_{k=a,b,c,d} (Y_2^k-3  C_{2S}^k)+ 5 \gt^2\bigg]\qquad  
\eea
for $f_0\to \infty$ and $\xi_{ab}\rightarrow -\frac16 \delta_{ab}$, where $Y_2^k$, $C_{2S}^k$ and $C_{2F}$ are defined in eq.~(\ref{Y2C2F}). We do not know of any previous determinations of the RGE in (\ref{eq:RGEY2}) and (\ref{eq:RGElambda2}).
We do not show  the RGE of the gauge couplings because they are not modified by the gravitational couplings (see the first paper in~\cite{ShapiroAnom} and~\cite{Narain:2012te,Narain,agravity}).

\subsubsection*{Anomalous generation of $1/f_0^2$}
However, the fact that $f_2$ and other gauge, Yukawa and quartic couplings 
start having non-vanishing $\beta$-functions means that the conformal-gravity computation becomes inconsistent
when going to higher orders.
The conformal mode of the agraviton, $\sigma$,  is a decoupled degree of freedom
in the classical Lagrangian of conformal gravity.  At quantum loop level,
$\sigma$  starts coupling to all terms that break scale invariance
proportionally to their $\beta$-functions,
so that $\sigma$ cannot no longer be gauged away.

Once $\sigma$ couples to other particles, they can propagate in loops within
Feynman diagrams containing, as external states,  $\sigma$ only.
This describes how the $R^2$ term is generated at a loop level high enough that the diagram contains running couplings.
The result can be written in terms of $\beta$-functions through the aid of consistency conditions obtained by formally 
promoting the couplings to fields, including the gravitational coupling.
A scalar quartic $\lambda$ starts contributing at  $\lambda^5$ order~\cite{Collins}; 
a gauge interaction starts contributing at $g^6$ order~\cite{Hathrell};
the effect of scalar quartics, Yukawa and gauge couplings was computed in~\cite{Osborn}
in parity-invariant theories.
The final result can be written as an RGE for $1/f_0^2$: 
\bea\label{eq:f0inv}
  \frac{d}{d\ln\mub}\frac{1}{f_0^2} &=& \frac{b_1 b_2 N_V}{18} \frac{g^6}{(4\pi)^8} + \frac{1}{25920(4\pi)^{12}} \left(6 \lambda_{abcd}\lambda_{cdmn}\lambda_{mnpq}\lambda_{aprs}\lambda_{bqrs}+ \right. \nonumber \\ && \left.+12 \lambda_{abcd}\lambda_{cdmn}\lambda_{mrpq}\lambda_{bspq}\lambda_{anrs} - \lambda_{acdm}\lambda_{bcdm}\lambda_{anrs}\lambda_{bnpq}\lambda_{rspq}\right)  + \cdots
\eea
in the limit $f_0\to \infty$ and $\xi_{ab}\rightarrow -  \delta_{ab}/6$.
We have written explicitly the leading gauge contribution assuming, for simplicity, a gauge group $G$
with a single gauge coupling  $g$, $N_V$ vectors and $N_f$ Weyl fermions 
in the same representation $R$ of $G$: $b_1$ and $b_2$ are the usual 
one-loop and two-loop $\beta$-function coefficients for $g$, precisely defined as
$  \sfrac{dg}{d\ln\mub} =  -b_1\sfrac{g^3}{(4\pi)^2} - b_2 \sfrac{g^5}{(4\pi)^4}  + \cdots$ 
and given by~\cite{Tarasov:1980au}\footnote{The group quantities $C_{2G}$, $C_{2F}$ and $T_F$ are defined 
as usual in terms of the generators $t^A$  in the representation $R$ as follows
\be [t^A, t^B] = i f^{ABC} t^C, \quad f^{ACD} f^{BCD}= C_{2G} \delta^{AB}, \quad t^At^A = C_{2F}, \quad  \mbox{Tr}(t^At^B) = T_F \delta^{AB}.   \ee
For example, for the vector representation of SU($N$) we have 
\be C_{2G} = N, \qquad  C_{2F} = \frac{N^2-1}{2N}, \qquad T_F =\frac12. \ee }
\be b_1 =  \frac{11}{3} C_{2G} -\frac23 T_FN_f, \qquad b_2 = \frac{34}3 C_{2G}^2 -\frac{10}3C_{2G}T_F N_f-2 C_{2F}T_FN_f . \ee
The gauge contribution to $1/f_0^2$ can  be either positive or negative depending on the field content.
For example, in the SM one has $N_V=3$, 
$b_1=19/6$ and $b_2=-35/6$ for $\SU(2)_L$ and $N_V=8$, $b_1=7$ and $b_2 =26$ for $\SU(3)_c$.
The  quartic of the Higgs doubled $H$, defined by the potential $ \lambda_H | H|^4$, contributes to the RGE for $1/f_0^2$ as
 $416 \lambda^5_H/5(4\pi)^{12}$,
  which is sub-dominant with respect to the gauge contributions.
Integrating the gauge contribution alone from infinite energy down to a scale where $g\ll 1$,
one finds $1/f_0^2 \simeq -  b_2 N_V g^4/72(4\pi)^6$.

\smallskip

The $\, \cdots\, $ in eq.\eq{f0inv} denote extra terms due to Yukawa couplings (partially computed in~\cite{Osborn})
and to gravitational terms (never computed and presumably first arising at order $f_2^6$).
The full unknown expression might perhaps take the form of a $\beta$-function of some combination of couplings,
given that the Weyl symmetry is not broken when all $\beta$-functions vanish.
Barring this exception, which seems not relevant (nature is neither described by a free theory nor by a conformal theory),
eq.\eq{f0inv} means that  conformal gravity is not a {complete}
 theory:
at some loop level, quantum corrections start generating back the extra couplings $f_0$ and $\xi_{ab}$ present in agravity.

 One important aspect of eq. (\ref{eq:f0inv}) is that its right-hand-side vanishes when all couplings sit at a fixed point, where all $\beta$-functions vanish. This tells us that the $f_0\to\infty$ limit is consistent when the other couplings on the right-hand-side approach a fixed point.



\subsubsection*{Anomalous generation of $\xi+1/6$}
Non-conformal $\xi$-couplings are generated at one-loop by the gravitational coupling $f_2$.
Starting from 
$\xi= -1/6$ at infinite energy,
$f_2$ induces a negative 
\be f_0^2(\xi+1/6) \sim - {\cal O}(f_2^2)\label{LambdaTildeGen}\ee
at finite energy.
However, as argued later, naturalness demands $f_2 \circa{<}10^{-8}$.
At perturbative level, $f_0$ alone
 does not generate $\xi+1/6$.
Extra anomalous contributions to $\xi+1/6$ are first generated at order $y^6/(4\pi)^6$,
$y^2\lambda^2/(4\pi)^6$, $\lambda^4/(4\pi)^8$
in the Yukawa couplings $y$ 
and in the scalar quartics $\lambda$ 
(see eqs.~(6.33) and (7.22) of~\cite{Osborn}, where individual terms have different signs; see also~\cite{Collins}).
For example the quartic couplings alone contribute as  
\beq \label{eq:RGEanomxi}
\frac{d\zeta_{ab}}{d\ln\mub}  =  \frac1{18(4\pi)^8} \left( \frac1{6} \lambda_{cpqr} \lambda_{dpqr}  \lambda_{cmna}\lambda_{dmnb}  + \lambda_{pqmn}\lambda_{pqcd} \lambda_{cmra} \lambda_{dnrb} - \lambda_{rpqd}\lambda_{rmnc} \lambda_{dmna} \lambda_{cpqb}  \right)  + \cdots
\eeq
for $f_0\to \infty$ and $\xi_{ab}\rightarrow -  \delta_{ab}/6$, where $\cdots$ denote the contribution of the other couplings.
In the SM Higgs case this contribution is
$ \sfrac{d\xi_H}{d\ln\mub} =48 \lambda^4_H/(4\pi)^8+\cdots$,
having written the potential as $\lambda_H |H|^4$ and the non-minimal coupling to gravity as $-\xi_H |H|^2 R$.

 It is  important to note that the right-hand-side of eq. (\ref{eq:RGEanomxi}) vanishes when all couplings sit at a fixed point, where all $\beta$-functions vanish. This tells us that the $f_0\to\infty$ limit is consistent when at the same time $\zeta_{ab}\rightarrow 0$ and the other couplings approach a fixed point. In this precise limit the conformal mode decouples from the rest of the degrees of freedom.
 

\section{The conformal mode of the agraviton}\label{sigma}
So far we have shown that a large self-coupling $f_0$ of the conformal mode of the agraviton
does not affect the rest of physics, provided that the non-minimal couplings $\xi$ of scalars go to the conformal value and the remaining couplings approach a fixed point.
We next address the big issue: 
what happens to the conformal mode of the agraviton when $f_0$ is big.


The one-loop agravity RGE for $f_0$, eq.\eq{RGEf0}, is valid for $f_0\ll 1$ and shows that a small $f_0$ grows with energy.
In general, when a dimensionless coupling behaves in this way, 
three qualitatively different things can happen depending on the non-perturbative
behaviour of the $\beta$-function
\beq \frac{df_0}{d\ln \mub} = \beta(f_0).\eeq
\begin{enumerate}
\item If $\beta(f_0)$ grows at large $f_0$  {faster than $f_0$}, then
$\int^\infty df_0/\beta(f_0)$ is finite and $f_0$ hits a Landau pole at finite energy.
The theory is inconsistent.\footnote{For example,
lattice simulations indicate that one scalar quartic or the gauge coupling in QED behave in this way~\cite{Frohlich:1982tw}.}

\item If $\beta(f_0)$ vanishes for some $f_0=f_0^*$, then $f_0$ grows up to $f_0^*$, entering into asymptotic safety.

\item If $\beta(f_0)$  remains positive but 
grows less than {or as} $f_0$, then $f_0$ grows up to $f_0=\infty$ at infinite energy.\footnote{For example, this behaviour is realised
if the $\beta$-function has the form  $\beta(f_0) = f_0 Z(f_0) $ with $Z(f_0) = b/(f_0^2 + f_0^{-2})/(4\pi)^2$ with $b>0$.
Then at low energy  $f_0$  runs logarithmically towards $f_0 \to 0$,
and  at large energy $1/f_0$ runs logarithmically towards $1/f_0 \to 0$.
Indeed, the full solution for $f_0^2>0$ is
$ f_0^{2} =t + \sqrt{1+t^2}$ \hbox{where} $ t =  b\ln( \sfrac{\mub}{\Lambda_0})/{(4\pi)^{2}}$
and $\Lambda_0$ is the transition scale at which $f_0\sim 1$.}
\end{enumerate}
In order to study what happens at large $f_0$, we can ignore all other couplings
and focus on the conformal mode of the agraviton.
We can choose a conformally flat background
$g_{\mu\nu}(x) = e^{2\sigma(x)} \eta_{\mu\nu}$,
as the background does not affect the UV properties of the theory.
Recalling eq.\eq{Lnotconfsigma}, the action for the conformal mode only is
\beq \label{eq:Ssigma}
S = \int d^4x \sqrt{|\det g|} \frac{R^2}{6f_0^2} =
\frac{6}{f_0^2}  \int d^4x (e^{-\sigma} \Box e^{\sigma})^2 =
 \frac{6}{f_0^2} \int d^4x  [\Box \sigma + (\partial \sigma)^2]^2 .
\eeq 
The field $\sigma$ has mass dimension 0, and its action in eq.~(\ref{eq:Ssigma}) respects the following symmetries:
shifts $\sigma(x)\to \sigma(x) + \delta\sigma$;
Poincar\'e invariance;
scale invariance;
{invariance under special conformal transformations:}
\beq \label{eq:confsigma}
\sigma(x)\to \sigma(x') - 2 c\cdot x,\qquad x'_\mu=x_\mu+ c_\mu x^2 - 2 x_\mu (c\cdot x),\eeq
at first order in the  infinitesimal constant vector $c_\mu$.
Conformal invariance here appears as a residual of the reparametrization invariance of the gravitational theory:
it is present because conformal transformations are those reparametrizations that leave
the metric invariant,
 up to an overall scale factor.
Being a residual of reparametrization invariance, this symmetry is non anomalous, up to the usual scale anomaly.
No other action is compatible with these symmetries.
Taking into account that $d^4x = (1+8c\cdot x) d^4 x'$,
the single terms in the action of eq.\eq{Ssigma} vary under a conformal transformation as
\begin{eqnsystem}{sys:conf}
\delta \int d^4x  \, (\partial\sigma)^4 &=& 8\int d^4x [ -  c\cdot \partial \sigma  (\partial\sigma)^2]\\
\delta \int d^4x \, (\partial\sigma)^2\Box \sigma &=&4 \int d^4x [ c\cdot \partial \sigma  (\partial\sigma)^2 -  c\cdot \partial \sigma \Box \sigma]\\
\delta  \int d^4x\,  (\Box\sigma)^2 &=&8  \int d^4x [ c\cdot \partial \sigma  \Box\sigma]  
\label{total derivative}
 \end{eqnsystem}
such that the combination in eq.~(\ref{eq:Ssigma})
 is invariant.\footnote{Alternatively, since conformal invariance can be seen as an inversion $x^\mu \to y^\mu = x^\mu/x^2$ followed
by a translation and by another inversion,
one can more simply check that the action is invariant under the inversion:
$d^4 x \to d^4 y/y^8$,
$ \sigma(x)\to \sigma(y) + \ln y^2$ and 
$[\Box_x\sigma + (\partial_x\sigma)^2]= y^4 [\Box_y\sigma + (\partial_y\sigma)^2]$. The transformation rule of $\sigma$ under the coordinate transformation $x^\mu \to y^\mu = x^\mu/x^2$ can be obtained by recalling its general definition in (\ref{sigmaIntro}) and that we are assuming here a conformally flat metric, i.e. $g_{\mu\nu}(x) = e^{2\sigma(x)} \eta_{\mu\nu}$.}
We verified, at tree-level, that the scattering amplitudes vanish, in agreement with the Coleman-Mandula theorem.

\medskip


For small $f_0$ one can compute the theory perturbatively around the 4-derivative kinetic term $(\Box\sigma)^2$.
As discussed in section~\ref{agravtysec}, this can be equivalently formulated as an SO(2)-invariant scalar $\Omega$ 
with a quartic coupling.
This shows that UV-divergent quantum corrections preserve the form of the action,
such that the quantum action is given by
\beq \Gamma = Z(f_0) \, S +\hbox{finite effects}.\eeq
Indeed, in the scalar theory with the field $\Omega$ and the simple quartic coupling all divergences can be reabsorbed by renormalising $f_0^2$ (which in that formulation represents the quartic coupling) and the field, $\Omega$. Going back to the formulation in terms of $\sigma$, both renormalisations (of $f_0$ and of $\Omega$) can be expressed in terms of a common rescaling of the action, which is what appears in eq.~(37). 

The common UV-divergent factor $Z(f_0)$ renormalises equally all terms in the action, such that it can be
seen as an RGE running of $f_0$, which we give here up to  two loops:
\beq \frac{d f_0^2}{d\ln\mub} =  \frac{1}{(4\pi)^2} \frac{5}{6} f_0^4  -  \frac{1}{(4\pi)^4} \frac{5}{12} f_0^6 +\cdots .\eeq 
The one-loop term reproduces the corresponding term in the full gravitational computation, eq.\eq{RGEf0}, while the two-loop term was never obtained before.
The Weyl anomaly, mentioned in section \ref{agravtysec}, affects this RGE only at higher loop level.
The reason is that the $\beta$-functions are already one-loop effects,
so that one needs at least two vertices and one extra loop to get a contribution from the anomaly. 
This remark not only applies to pure anomalous effects, but also to mixed $f_0$-anomaly contributions; in the latter case, indeed, a couple of internal $\sigma$-lines should be converted to the particles which $\sigma$ couples  to through the anomaly and again at least two vertices proportional to $\beta$-functions and one extra loop are needed.

\medskip

When $f_0$ grows the path-integral receives contributions from fluctuations of $\sigma$ 
with larger and larger amplitude, probing  the  terms in the action of eq.\eq{Ssigma} with higher powers in $\sigma$.
For large $f_0$ the action becomes dominated by
the $(\partial\sigma)^4$ term that has the highest power of $\sigma$, while the kinetic term becomes negligible.
This can happen because all terms in the action have the same number of derivatives.
For example, a field configuration $\sigma(r) = \sigma_0 e^{-r^2/a^2}$ contributes
as
$S \sim (\sigma_0 + \sigma_0^2)^2/f_0^2$, independently of the scale $a$,
such that for $f_0\circa{>}1$ the path integral is dominated by the second term.

In the limit $f_0\to \infty$ the action $S$ simplifies to
\beq S_\infty = \frac{6}{f_0^2}\int d^4x\, (\partial\sigma)^4. \label{Sinfty}\eeq
Although for large $f_0$ the theory is non-perturbative in $f_0$, one can still develop an analytical argument to show the absence of a Landau pole of $f_0$, as we now discuss. 
The action in eq.~(\ref{Sinfty}) acquires new symmetries: $S_\infty$  is $\Z_2$-invariant ($\Z_4$-invariant if complexified);
furthermore,
being the term of $S $ with the highest  power of $\sigma$,
it is invariant under the homogeneous part of the transformation in eq.\eq{confsigma}, while the other two terms, 
$ (\partial\sigma)^2\Box \sigma$ and $(\Box \sigma)^2$ or any combination of them, are not.
Symmetries imply that the quantum action $\Gamma_\infty$, which includes the classical and UV-divergent quantum corrections, is fully described by $\Gamma_\infty = Z_\infty S_\infty$,
where $Z_\infty$ is a constant, related to the $Z(f_0)$ in the full theory as $Z_\infty=\lim_{f_0\to\infty}Z(f_0)$. 
This constant must equal unity, $Z_\infty=1$ because the theory is classical at
large field values, for which  $S_\infty \gg 1$, and because its form at all field values is fixed by symmetries.
The theory with action $S_\infty$, despite  being interacting, behaves as a free theory,
in the sense that the quantum action does not receive 
divergent corrections.

This shows that, in the full theory, $f_0$ can flow to large values without hitting  Landau poles:
$\beta(f_0) ={\cal O}(1/f_0)$ at $f_0 \gg 1$.
Having distilled the non-perturbative dynamics of the conformal mode of the agraviton in a simple action, eq.\eq{Ssigma},
it seems now feasible to fully clarify its dynamics.
We have shown that it hits no Landau poles, excluding case 1.\ of the initial list.
The theory at $f_0 \gg 1$ should be computable by developing a perturbation theory in $1/f_0$.
We have not been able of excluding case 2: a vanishing $\beta(f_0)$ at $f_0 \sim 4\pi$.
Non-perturbative numerical techniques seem needed to determine the behaviour of the theory at the intermediate
energy at which $f_0 \sim 4\pi$,
although this currently needs adding a regulator that breaks the symmetries of the theory (such as a lattice or a momentum 
averager~\cite{wetterich}), obscuring possible general properties (such as the sign of $\beta(f_0)$) that
could follow from the positivity of the symmetric action in eq.\eq{Ssigma}.

\bigskip

The letter  `{\em a}' in the name `conformal mode of the {\em a}graviton' reminds 
that our field $\sigma$ contains two degrees of freedom because its action contains 4 derivatives,
while the usual `conformal mode of the graviton' obtained from the Einstein action only contains one degree of freedom.
More precisely, the Einstein term alone,
$-\frac12 \bar M_{\rm Pl}^2R $, where $\bp$ is the reduced Planck mass,
gives a negative kinetic term 
$ 3\bar M_{\rm Pl}^2  \Omega_L \Box \Omega_L$
for $\Omega_L =e^\sigma$, see eq.\eq{dR}.
Summing the Einstein term with  $R^2/6 f_0^2$,
the 4-derivative conformal mode of the agraviton $\sigma$ splits into a physical mode with 
positive kinetic term and mass $M_0 = f_0 \bar M_{\rm Pl}/\sqrt{2}$ for $f_0\ll1$,
and the usual massless Einstein term, which is reparametrization-dependent.\footnote{Many authors refuse to view the theory with higher derivative as legitimate because of the consequent ghosts:
see e.g.~\cite{0811.2197} for attempts to discard the $(\Box\sigma)^2$ term.
Accepting the presence of higher derivatives allows to describe the
Weyl anomaly as ordinary RGE running of $f_{0,2}$, rather than 
by modifying Einstein gravity by adding a complicated `quantum anomalous action'~\cite{Antoniadis}
which encodes the anomalous behaviour of generic undefined theories of gravity.}
To see this, it is convenient to use the form of the action where  $\sigma$ is rewritten in terms of
two fields with two derivatives, $\Omega_L$ and $\Omega_R$, (see section \ref{agravtysec}).
Adding to the previous discussion the Planck mass 
the  Lagrangian becomes
 \beq\label{eq:notSO11}
 \Lag = -2\Omega_R \Box\Omega_L - \frac16 f_0^2 \Omega_L^2 \Omega_R^2
+ 3 \bp^2 \Omega_L \Box\Omega_L 
.\eeq
We expand in fluctuations around the minimum, $\Omega_R =0$ and  $\Omega_L=1$, where 
we arbitrarily choose unity in order to keep the metric as $\eta_{\mu\nu}$, 
while other values would correspond to a different unit of mass.
Then, the quadratic part of the action can be diagonalized by defining 
$\Omega_L = 1 + (\alpha+\beta)/\sqrt{3}\bp$, $\Omega_R = \sqrt{3} \bp \beta$,
where $\alpha$ is the Einstein ghost and $\beta$ is the massive scalar component of the graviton.
The result is 
\beq\label{eq:splittedconf}
 \Lag = \alpha \Box \alpha + \beta (-\Box - M_0^2)\beta -V\qquad\hbox{with}\qquad
V = \frac16 f_0^2 \beta^2 (\alpha+\beta) (\alpha+\beta+2\sqrt{3}\bp).\eeq


\begin{figure}[t]
\begin{center}
\includegraphics[width=0.45\textwidth]{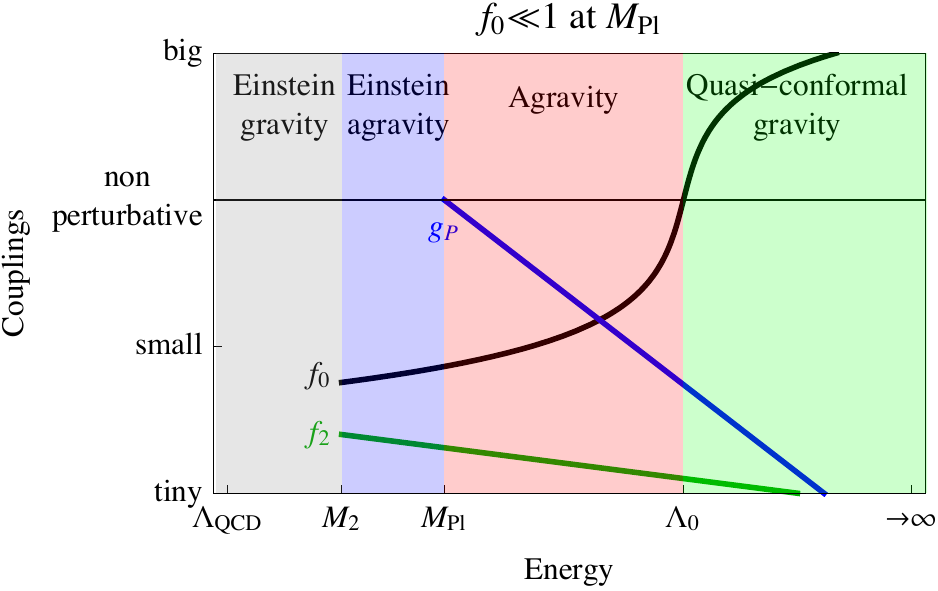}\qquad
\includegraphics[width=0.45\textwidth]{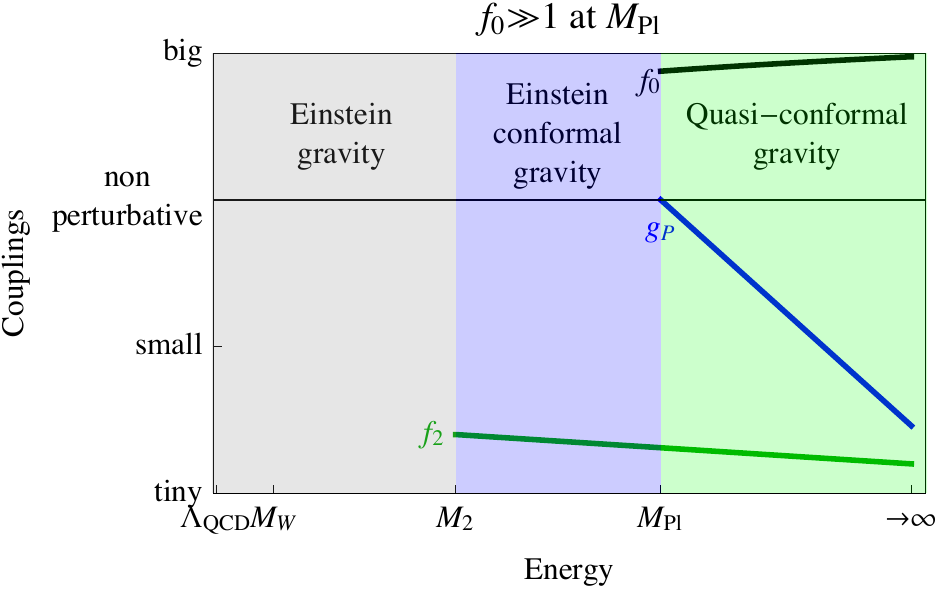}
\caption{\em RGE running of the main dimensionless couplings $f_0, f_2, g_P$ in the two possible scenarios
that do not lead to unnaturally large corrections to the Higgs mass:
$f_0 , f_2 \ll 1$ at the Planck scale (left),
$f_2 \ll1$ and $ f_0 \gg 1$ at the Planck scale (right). Here 
$M_{\rm Pl}$ is the Planck mass,
$M_2\equiv f_2 \bp/\sqrt{2}$ is the graviton ghost mass,
$\Lambda_{0}$ is the RGE scales at which $f_{0}\sim 4\pi$.
\label{fig:RGEAgravity}}
\end{center}
\end{figure}

\section{Scenarios compatible with naturalness of the Higgs mass}\label{nat}
In the following we discuss implications of case 3.
Qualitatively different scenarios can arise, depending on the ordering between the key scales:
\begin{itemize}
\item $\Lambda_0$, the energy scale at which the self-coupling of the conformal mode equals $f_0 \sim 4\pi$,
with $f_0 \ll 4\pi $  at $E \ll \Lambda_0$ and $f_0 \gg 4\pi$ at $E\gg \Lambda_0$.

\item $\Lambda_2$, the energy scale at which the graviton self-coupling equals $f_2 \sim 4\pi$,
with $f_2 \ll 4\pi $  at $E \gg \Lambda_2$.

\item The Planck scale. 
As this is the largest known mass scale, in the context of dimensionless theories
it can be interpreted as the largest dynamically generated vacuum expectation value or condensate.
\end{itemize}
The scales $\Lambda_{0,2}$ can be physically realised in nature (like the scale $\Lambda_{\rm QCD}$ at which $\SU(3)_c$ becomes strong)
if they are larger than the Planck scale. Otherwise they are not realized 
(like the scale at which $\SU(2)_L$ would have become strong, if symmetry breaking had not occurred at a higher energy)
and we use $\Lambda_{2}\ll M_{\rm Pl}$ to denote $f_{2} \ll 1$ at $M_{\rm Pl}$ where $M_{\rm Pl}$ is the Planck mass.

\smallskip

In this section we adopt Higgs mass naturalness as a criterion to limit the possible speculations.
For example, the simplest possibility in which the Planck scale is identified with $\Lambda_2$ or $\Lambda_0$
leads to unnaturally large physical corrections to the Higgs mass from gravity.
Naturalness demands $f_2 \ll 1$ at the Planck scale, while $f_0$ can be either very small or very large,
giving rise to two natural possibilities shown in fig.\fig{RGEAgravity}:
$f_0\ll 1$ at $M_{\rm Pl}$ (left panel) and $f_0 \gg 1$ at $M_{\rm Pl}$ (right).


\subsection{$f_0 \ll 1$ at the Planck scale}
The first possibility is the one considered in~\cite{agravity}, that
showed that the Planck mass can be dynamically generated, within a dimensionless theory, 
from a dynamically induced vacuum expectation value  of a fundamental scalar $S = (s+i s')/\sqrt{2}$. The part of the dimensionless Lagrangian involving $S$ and the
SM Higgs doublet $H$  is
\beq \Lag =    \bigg[
 |D_\mu S|^2 - \lambda_S |S|^4   -\xi_S |S|^2 R\bigg]+
\bigg[  |D_\mu H|^2  -\lambda_H |H|^4 - \xi_H |H|^2 R \bigg]+ \lambda_{HS}|S|^2 |H|^2
 .\eeq
Provided that $\lambda_S$ runs in such a way that it vanishes at the same scale at which its $\beta$-function vanishes,
$s$ gets a vacuum expectation value with cosmological constant tuned to zero, and
$\bar M_{\rm Pl}^2 = \xi_S \langle s\rangle^2$ is positive provided that the parameter $\xi_S$, renormalized at the Planck scale, is positive.
An unpleasant feature of the model is that the mixed quartic $\lambda_{HS}$ must be very small, in order to avoid 
inducing an unnaturally large contribution to the Higgs mass
($M_h^2 = \lambda_{HS} \langle s\rangle^2$, that appears in the potential as $- M_h^2 |H|^2/2$).
Refs.~\cite{agravity,1502.01334} showed that $\lambda_{HS}$ can be naturally small, despite being generated at loop level through gravity loops as
\beq \label{eq:RGElambdaHS}(4\pi)^2 \frac{d{\lambda_{HS}}}{d\ln\mub} = -\xi_H\xi_S [ 5\gt^4  + 36 \tilde\lambda_H \tilde\lambda_S] + \cdots \qquad \hbox{($f_0 \ll 1$)},\eeq
where $\tilde\lambda_{S} \equiv f_0^2 (\xi_S+1/6)$ and $\tilde\lambda_H \equiv f_0^2(\xi_H+1/6)$ are the
couplings that appear in the perturbatively equivalent formulation of agravity of eq.\eq{Snotconf2}, where
$f_0$ and $\xi_{H,S}$ become quartic couplings with an extra scalar $\vec\Omega$. 
The Higgs mass is natural if $f_{0,2} \circa{<}10^{-8}$.
The above scenario needs to be reconsidered:
\begin{enumerate}
\item[a)] Is naturalness still satisfied, or $f_0$ becoming strongly coupled at the energy scale $\Lambda_0$
generates a $\tilde\lambda_{H,S}$ of the same order?
\item[b)] Can one get $\xi_S > 0$ at the Planck scale starting from $\xi_S = -1/6$ at infinite energy?
\end{enumerate}
A peculiar RG running behaviour at a very large scale, such as $\Lambda_0\circa{>} 10^{10^{16}}\GeV$,
does not imply perturbative contributions to scalar masses of the same order, 
as long as no new physics nor vacuum expectation values nor
condensates develop at that scale~\cite{1701.01453}.
Non-perturbative ultra-Planckian
contributions to the cosmological constant and the Planck mass from a $f_0\sim 4\pi$
are forbidden by the global shift symmetry  $\sigma \to \sigma +  \delta\sigma$.
Planckian corrections to the cosmological constant remain unnaturally large as usual.

The answer to a) seems positive: as shown in section~\ref{agravtysec}
perturbative corrections in $f_0$ behave like quartic scalar couplings,
and thereby renormalise the $\tilde\lambda_{H,S}$ couplings
(mixed quartics between the scalars and the conformal mode of the graviton)
 only multiplicatively, like in the one-loop RGE, eq.\eq{RGExi}.
The same happens at $f_0 \gg 1$:
non-vanishing $\tilde\lambda_{H,S}$ are only generated 
by $f_2$ (see eq.~(\ref{LambdaTildeGen})) and by the multiloop anomalous effects discussed in section \ref{conformal}.
Non-perturbative corrections in $f_0 \sim 4\pi$ presumably too renormalise $\tilde\lambda_{H,S}$  only multiplicatively,
as the scalars $H,S$ are not involved in the strong self-coupling of the conformal mode of the graviton.

Concerning issue b), the answer can be positive in a theory where $\xi_S$ is very close to $-1/6$ around and above the energy scale $\Lambda_0$,
and a positive $\xi_S$ is only generated trough anomalous running (see e.g.\ eq.\eq{RGEanomxi}) at a much lower energy
where $f_0 \ll 1$ by some matter coupling becoming non-perturbative.

\medskip

Given that non-perturbative physics seems anyhow necessary, we propose here a simpler mechanism 
for the generation of the Planck mass that relies on a new strong coupling $g_P$, rather than on a perturbative coupling $\lambda_S$.
Without introducing any extra scalar $S$ (and thereby bypassing the issue of a small $\lambda_{HS}$),
the Planck scale can be induced by a new gauge group $G$ (under which the Higgs is neutral) 
with a gauge coupling $g_P$ that runs to non-perturbative values around the Planck scale, such that condensates $f$ are generated.
This is shown as blue curve in fig.\fig{RGEAgravity}.
This scenario can be very predictive, as one coupling $g_P$ dominates the dynamics.
The sign of  $M_{\rm Pl}^2$ is predicted: however, it is not determined by dispersion relations and seems to depend
on the detailed strong dynamics of the model (gauge group, extra matter representations)~\cite{Sakharov}.

One has the desired $M_{\rm Pl}^2>0$ provided that the theory admits an effective-theory approximation where
the effect of the strong dynamics is dominantly encoded in a mixing of the graviton with a composite spin-2 resonance,
analogously to how a photon/$\rho$ mixing approximates QCD effects.
Then, the relevant effective Lagrangian for the graviton $h_{\mu\nu}$ and the spin-2 resonance is
\beq \Lag_{\rm eff} = - \frac{M^2}{2} R_\rho +f^4[ a (h_{\mu\nu}-\rho_{\mu\nu})^2 + (h_{\mu}^{\,\, \,\mu} - \rho_{\mu}^{\,\, \,\mu})^2] 
+{\cal O}(\partial^4 h_{\mu\nu})+{\cal O}(\partial^4 \rho_{\mu\nu}).\eeq
The first term 
is the positive quadratic kinetic energy for the spin-2 resonance generated by strong dynamics;
we wrote it as a `curvature' $R_\rho$ multiplied by some positive $M^2>0$.
The second term is a mass term, which presumably approximatively has Fierz-Pauli form, $a\approx 1$.\footnote{It can be rewritten in a covariant form as the mass term resulting, in the unitary gauge, from the spontaneous symmetry breaking 
of general coordinate invariance acting separately on ordinary fields and on composite fields, 
$\hbox{GL}_h \otimes \hbox{GL}_\rho \stackrel{f}\to \hbox{GL}$~\cite{Nima}.}
Next, we integrate out $\rho_{\mu\nu}$ obtaining an effective action for the graviton $h_{\mu\nu}$.
At leading order in derivatives one simply has $\rho_{\mu\nu}=h_{\mu\nu}$, irrespectively of the precise form of the mass term.
Thereby the resulting effective action is the Einstein action, with $\bar M_{\rm Pl}^2=M^2$.

\smallskip

Furthermore, the strong dynamics generates at the same time a cosmological constant.
In a theory with no matter charged under $G$ it is negative and of order $M_{\rm Pl}^4$:
\beq
V=\frac{T_{\mu}^{\,\, \,\mu}}{4}= \frac{\partial_\mu \mathscr{D}^\mu }{4}=\frac{1}{4} \frac{\beta_{g_P}}{2g_P}  \langle F_{\alpha\beta}^{A2}\rangle  
\eeq
where $\mathscr{D}_\mu $ is the anomalous dilatation current and $\beta_{g_P} <0$.


This  large contribution to the cosmological constant can be avoided if the theory also
includes a Weyl fermion $\lambda$ in the adjoint of the gauge group $G$, such that the
most general dimensionless action
\beq {S} = \int d^4x \sqrt{|\det g|} \bigg[ - \frac{F_{\mu\nu}^{A2}}{4 g_P^2}  +  \bar \lambda^A  i\slashed{D} \lambda^A+
\frac{R^2}{6f_0^2} + \frac{\frac13 R^2 -  R_{\mu\nu}^2}{f_2^2} 
\bigg]
\eeq
 is accidentally supersymmetric in its strongly coupled sector.
With this particle content  $\langle F_{\alpha\beta}^{A2}  \rangle=0$ vanishes, being the $D$-term of
an accidental unbroken global  supersymmetry, while the fermion condensate can be computed~\cite{SUSY}.

The Higgs has no renormalizable interaction with the strong sector at the Planck scale:
it is only generated through gravitational loops, 
between the Planck mass and the masses $M_{0,2}$ of the extra components of the agraviton.
The one-loop RGE for the Higgs mass in this regime was computed in~\cite{agravity},
and the contribution proportional to $\bp^2$ is
\beq
(4\pi)^2\frac{d  }{d\ln\mub} {M_h^2}= -\xi_H [5\gt^4+\gs^4(1+6\xi_H)] \bar M_{\rm Pl}^2 + \cdots\qquad \hbox{for}\qquad M_{0,2} < \bar\mu < M_{\rm Pl}
\eeq
%
{where $\cdots$ are contributions that are not dangerous from the point of view of naturalness}. In appendix~\ref{RGEM} we write the
one-loop RGE for the most general massive parameters.

%
%
%
%
%
%


\subsection{$f_0\gg 1$ at the Planck scale}
A simpler alternative that avoids having a very large RGE scale at which $f_0$ crosses  $4\pi$ 
is that $f_0$ is still large at the Planck scale and never gets small.

The conformal mode of the agraviton only has small anomalous couplings, 
until its dynamics suddenly changes 
when 
some vacuum expectation value or condensate is first generated.
We assume that the largest such effect is the Planck mass, that can be generated
in the ways discussed in the previous section.
Then, the tree-level Lagrangian of eq.\eq{splittedconf} describes how $\sigma$ splits into two-derivative modes.
The SO(1,1) symmetry that prevented quantum corrections to the strongly-interacting theory with $f_0\gg 1$ gets broken by $M_{\rm Pl}$.

The physical difference with respect to the previous case is that
only the Einstein conformal mode of the graviton appears
in the effective theory below the Planck scale down to the scale $M_2$.
The RGE are those of gauge-fixed conformal gravity {(see eqs. (\ref{eq:RGEf2c}), (\ref{eq:RGEY2}) and (\ref{eq:RGElambda2}))}.  {Proceeding as in appendix~\ref{RGEM},  the RGE of the Higgs mass is
\beq
(4\pi)^2\frac{d  }{d\ln\mub} {M_h^2}= \frac{5}{6} \gt^4 \bar M_{\rm Pl}^2 + \cdots,\qquad \hbox{for}\qquad M_2 < \bar\mu < M_{\rm Pl},
\eeq
which is naturally small for $f_2 \circa{<}10^{-8}$.

%
%


\section{Conclusions}\label{concl}
In dimensionless gravity  theories (agravity), the conformal mode of the agraviton 
consists of two fields: the usual conformal mode of the graviton and an extra scalar, 
jointly described by a 4-derivative action for a single field $\sigma$, defined by $g_{\mu\nu}(x) =e^{2\sigma(x)} \eta_{\mu\nu}$.
The self-interactions of the conformal mode of the agraviton are controlled by a  coupling $f_0$ that is not asymptotically free.
In section~\ref{agravtysec} we recomputed its RGE, and extended it at the two-loop level, by developing a formulation where
$f_0^2$ becomes an extra scalar quartic coupling.
In the presence of scalars, their  dimensionless $\xi$-couplings to gravity
become scalar quartics, and the whole agravity can be rewritten
as conformal gravity plus two extra scalars with an SO(1,1) symmetry.
This perturbative equivalence allowed us to recompute the one-loop RGE equations of a generic agravity theory,
confirming previous results~\cite{agravity}, writing them in an equivalent simpler form
where no couplings appear at the denominator in the $\beta$-functions,
extending them at two loops.

\smallskip

In particular, rewriting $f_0^2$ as a quartic scalar clarifies why a small $f_0$ grows with energy in any agravity theory.
A Landau pole would imply that agravity is only an effective theory and that the Higgs mass receives unnaturally large corrections.

\smallskip

In section~\ref{agravtysec}, \ref{conformal} and \ref{sigma} we have shown that, nevertheless, agravity can be a complete theory.
Agravity can be extrapolated up to infinite energy, although in an unusual way:
the dimensionless coupling $f_0$ grows with energy, 
becomes strongly coupled above some critical RGE scale $ \Lambda_0$,
and can smoothly grow to $f_0\to \infty$ at infinite energy. Although we have excluded that $f_0$ has a Landau pole, i.e. that it blows up at finite energy, there is another possibility which we have not studied in the present work: $f_0$ can approach asymptotically a finite non-perturbative fixed point. Analysing this possibility requires having control on intermediate regimes where $f_0 \sim 4 \pi$, which is beyond our current ability.

Provided that  all scalars are asymptotically conformally coupled (all $\xi$-couplings must run approaching $-1/6$) and all matter couplings approach a fixed point (possibly a free one, like in QCD)  in the UV, the simultaneous $f_0\to \infty$ limit turned out to be consistent. In this case and
in the limit of infinite energy  the conformal mode of the agraviton fluctuates freely and  decouples from the rest of the theory.
In the UV limit the theory can then be computed by viewing $\sigma$ as a gauge redundancy, that can be fixed with the Faddeev-Popov procedure.
 One then obtains conformal gravity at infinite energy. In section~\ref{conformal} we provided the one-loop RGE at the zero order in the expansion in $1/f_0^2$ and $\xi+1/6$, including  the most general matter sector.

However, the conformal symmetry is anomalous and its violation is dictated by renormalization group equations that describe how
the dimensionless parameters that break conformal symmetry, $f_0$ and $\xi+1/6$, are generated at a  few-loop order.
As a result, at energies much above $\Lambda_0$
the conformal mode of the agraviton $\sigma$ is strongly self-coupled ($f_0\gg 1$)
and fluctuates wildly, being negligibly coupled to other particles.
In section~\ref{sigma} we isolated its peculiar action and showed that, despite the strong coupling, it can be controlled through its symmetries.
The action is simple enough that its full quantum behaviour could be simulated on a Euclidean lattice.




The anomalous multi-loop RGE which generate $1/f_0^2$ and $\xi+1/6$, are not (yet) fully known, but it is already possible to discuss the physical implications of this theory.
We assume that the largest mass scale dynamically generated through vacuum expectation values or condensates is the Planck scale.
Two situations discussed in section~\ref{nat} can lead to a scenario where the Higgs mass does not receive unnaturally large corrections.
If $f_0 \ll 1$ at the Planck scale
 one obtains agravity at sub-Planckian energies: 
we wrote the most general RGE for massive parameters, and argued that a new gauge group with a fermion in the adjoint
can become strongly coupled around the Planck scale and successfully generate $\bp$, without generating a
Planckian cosmological constant (this mechanism was never explored before in the context of agravity).
Alternatively,  $f_0 \gg 1$ at the Planck scale seems  a viable possibility:
in this case the scalar component of the agraviton is above the Planck scale.


\smallskip

\appendix

\section{One-loop RGE in agravity}\label{RGE}
When $f_0$ and all couplings are small, the one-loop $\beta$-functions $\beta_p \equiv dp/d\ln\mub$ of all parameters $p$ of
the generic agravity theory of eq.\eq{Lconf} and eq.\eq{Lnotconf},
can be conveniently written in terms of  the 
combination of parameters that appear in eq.\eq{Snotconf2},
$\zeta_{ab} = \xi_{ab}+\delta_{ab}/6$ and
\beq\label{eq:tlambda}
\tilde\lambda_{abcd} = \lambda_{abcd} + 3f_0^2 (\zeta_{ab} \zeta_{cd}+
\zeta_{ac} \zeta_{bd}+\zeta_{ad} \zeta_{bc}),\qquad
\tilde\lambda_{ab} = f_0^2 \zeta_{ab},\qquad\tilde\lambda = f_0^2.\eeq
The RGE are
\begin{eqnsystem}{sys:agravityRGE2}
(4\pi)^2\frac{d \gt^2}{d\ln\mub}&=& -\gt^4\bigg(\frac{133}{10} +\frac{N_V}{5}+\frac{N_f}{20}+\frac{N_s}{60}\bigg), \label{RGEf2}\\
\riga{where $N_V$, $N_f$, $N_s$ are the number of vectors, Weyl fermions and real scalars (in the SM $N_V=12$, $N_f = 45$, $N_s = 4$), and 
}\\
(4\pi)^2 \frac{df_0^2}{d\ln\mub} &=& \frac53 f_2^4+5f_2^2 f_0^2+ \frac56 f_0^4  + 3 \tilde\lambda_{ab}\tilde\lambda_{ab},\\
(4\pi)^2 \frac{d\tilde\lambda_{abcd}}{d\ln\mub} &=&  \sum_{\rm perms} \bigg[\frac18
  \tilde\lambda_{abef}\tilde\lambda_{efcd}+
  \frac38 \{\theta^A,\theta^B\}_{ab}\{\theta^A ,\theta^B\}_{cd}
  -\Tr\, Y^a Y^{\dagger b} Y^c Y^{\dagger d}+\label{eq:RGElambda}
\\
&&
+\frac{5}{288} f_2^4 \delta_{ab}\delta_{cd}+ 
\frac1{16}\tilde\lambda_{ab}  \tilde\lambda_{cd}\bigg]
+ \tilde\lambda_{abcd} \bigg[5f_2^2 + \sum_{k=a,b,c,d} (Y_2^k-3  C_{2S}^k)\bigg]   ,\nonumber 
\\
(4\pi)^2 \frac{d\tilde\lambda_{ab}}{d\ln\mub} &=& \frac{f_0^2\tilde\lambda_{ab}}3 +
\tilde\lambda_{abcd}\tilde\lambda_{cd} +5f_2^2\tilde\lambda_{ab}+  \tilde\lambda_{ab} \sum_{k=a,b} \left(Y_2^k - 3C_{2S}^k  \right)  
+2\tilde\lambda_{ac}\tilde\lambda_{cb}  + \frac{5 f_2^4\delta_{ab}}{18},\label{eq:RGExi}
\\
\label{eq:RGEY}
(4\pi)^2 \frac{dY^a}{d\ln\mub} &=& \frac{Y^{\dagger b}Y^b Y^a + Y^a Y^{\dagger b}Y^b}{2}+ 2 Y^b Y^{\dagger a} Y^b +
 Y^b \Tr(Y^{\dagger b} Y^a)\nonumber  - 3 \{ C_{2F} , Y^a\}  + \frac{15}{8}\gt^2 Y^a.
 \end{eqnsystem}
 The sum over ``perms" runs over the $4!$ permutations of $abcd$ and $Y_2^k$, $C_{2S}^k$ and $C_{2F}$ are defined  by
\be {\rm Tr} (Y^{\dagger a}Y^b) =Y_2^a \delta ^{ab}, \qquad \theta^A_{ac} \theta^A_{cb}= C_{2S}^{a} \delta_{ab},\qquad C_{2F} = t^A t^A, \label{Y2C2F}\ee 
where $\theta^A$ and $t^A$ are the generators of the gauge group for scalars and fermions respectively {(the gauge couplings are contained in $\theta^A$ and $t^A$)}.


\section{One-loop RGE for massive parameters in agravity}\label{RGEM}

For the sake of completeness we also write the RGE for  the most generic massive parameters
that can be added while keeping the theory renormalizable:
the reduced Planck mass $\bp=M_{\rm Pl}/8\pi$, the cosmological constant $\Lambda$,  scalar squared masses $m^2_{ab}$,
 scalar cubics $A_{abc}$,  fermion masses $M_{ij}$ defined as
\beq \label{massiveL}\Lag_{\rm massive} =-   \frac12 \bp^2 R - \Lambda-  \frac12 m^2_{ab} \phi_a \phi_b - \frac16 A_{abc}\phi_a\phi_b\phi_c
-\frac12 (M_{ij}\psi_i\psi_j+\hbox{h.c.}). 
\eeq

The RGE for the massive terms can be obtained from the generic dimensionless RGE 
by considering one neutral scalar $s$ as a dummy non-dynamical variable, such that
\beq
\bp^2 = \xi_{ss} s^2 ,\qquad
\Lambda =\lambda_{ssss}  \frac{s^4}{4!} ,\qquad
M_{ij} =Y^s_{ij} s,\qquad m^2_{ab} = \lambda_{abss}\frac{s^2}{2},\qquad
A_{abc}=\lambda_{abcs}s.
\eeq
This technique has been used to determine the RGE of massive parameters in generic QFT without gravity \cite{Luo:2002ti}. 
Gravitational couplings have been included in some less general models in~\cite{Avramidi:1986mj,Shapiro:2000dz}.
The generic RGE of massive parameters in agravity are
\begin{eqnsystem}{sys:RGm} 
(4\pi)^2\frac{d  \bp^2}{d\ln\mub}  &=& \frac{1}{3} m_{aa}^2+\frac{1}{3}\Tr(M^\dagger M) +2\xi_{ab} m_{ab}^2  +\left(\frac{2f_0^2}{3}-\frac{5f_2^4}{3f_0^4}+2X\right)\bp^2,\label{RGEPlanck}\\
(4\pi)^2\frac{d  \Lambda}{d\ln\mub}  &=&\frac{m_{ab}^2 m_{ab}^2}{2}-
\Tr[(MM^\dagger)^2]+\frac{5f_2^4+f_0^4}{8} \bp^4+(5f_2^2+f_0^2) \Lambda+4\Lambda X,\\
 (4\pi)^2 \frac{dM}{d\ln\mub}  &=& \frac12(Y^{\dagger b}Y^b M + M Y^{\dagger b}Y^b)+ 2Y^b M^{\dagger } Y^b +  Y^b \Tr(Y^{\dagger b} M)+ \nonumber\\
&& - 3 \{ C_{2F} , M\}  + \frac{15}{8}f_2^2 M+MX,\\
 (4\pi)^2\frac{d m^2_{ab}  }{d\ln\mub} &=&\lambda_{abef} m_{ef}^2+ A_{aef}A_{bef} -2
  \big[\Tr(Y^{\{a} Y^{\dagger b\}} M M^{\dagger})+\nonumber \\ && 
 +\Tr( Y^{\dagger \{a} Y^{b\}}  M^\dagger M)
+\Tr\, (Y^a M^{\dagger }Y^b M^{\dagger})+  \Tr\, (MY^{\dagger a} MY^{\dagger b} )\big]+\nonumber\\ && \nonumber+\frac52 f_2^4 \xi_{ab} \bp^2+\frac{f_0^4}{2}\left(\xi_{ab}+6\xi_{ae}\xi_{eb}\right) \bp^2+ \\
 &&+f_0^2 \left(m_{ab}^2 +3\xi_{bf} m_{af}^2+3\xi_{af}m_{bf}^2 +6 \xi_{ae}\xi_{bf} m_{ef}^2\right)+\nonumber\\ &&+m_{ab}^2 \left[\sum_{k=a,b} (Y_2^k-3  C_{2S}^k)+ 5 f_2^2+2X\right], \\
  (4\pi)^2\frac{d A_{abc}  }{d\ln\mub} &=& \lambda_{abef}A_{efc}+ \lambda_{acef}A_{efb}+ \lambda_{bcef}A_{efa}+ \nonumber \\
&&   -2\Tr\left(Y^{\{a}Y^{\dagger b} Y^{c\}}M^\dagger\right)- 2\Tr\left(Y^{\dagger\{c}Y^{a} Y^{\dagger b\}}M\right)+\nonumber \\&&+f_0^2\left(A_{abc}+3\xi_{af}A_{fbc}+3\xi_{bf}A_{fac}+3\xi_{cf}A_{fab}\right)+ \nonumber\\&&+6f_0^2\left(\xi_{ae}\xi_{bf}A_{efc}+\xi_{ae}\xi_{cf}A_{efb}+\xi_{be}\xi_{cf}A_{efa}\right)+\nonumber \\  && + A_{abc}\left[\sum_{k=a,b,c} (Y_2^k-3  C_{2S}^k)+ 5 f_2^2+X\right],  \label{RGEcubic}
\end{eqnsystem}
where the curly brackets represent the sum over the permutations of the corresponding indices: e.g. $Y^{\{a} Y^{\dagger b\}}= Y^{a} Y^{\dagger b}+Y^{b} Y^{\dagger a}$.
Notice that $X$ is a gauge-dependent quantity, equal to
\beq X= \frac{(3c_g^2-12c_g+13)\xi_g}{4(c_g-2)}+\frac{3(c_g-1)^2f_0^2}{4(c_g-2)^2}\eeq
using the gauge-fixing action of~\cite{agravity}, which depends on two free parameters $\xi_g$ and $c_g$:
\beq \label{eq:gf}
S_{\rm gf} = - \frac{1}{2\xi_g}\int d^4x ~ f_\mu \partial^2 f_\mu,\qquad
f_\mu = \partial_\nu( h_{\mu\nu} - c_g \frac12 \eta_{\mu\nu} h_{\alpha \alpha}).
\eeq
The RGEs of massive parameters are gauge dependent because the unit of mass is gauge dependent.
Any dimensionless ratio of dimensionful parameters is physical and the corresponding RGE is gauge-independent,
as it can be easily checked from eqs. (\ref{RGEPlanck})-(\ref{RGEcubic}).
For example~\cite{agravity} gave the RG equation for $M_h^2/\bp^2$.


\small

\subsubsection*{Acknowledgments}
We thank Andrei O. Barvinsky, Agostino Patella, Alberto Ramos, Francesco Sannino, Ilya  L.\  Shapiro,
 Andreas Stergiou, Nikolaos Tetradis, Enrico Trincherini and Hardi Veermae for useful discussions.
This work was supported by the ERC grant NEO-NAT.


\footnotesize



\begin{thebibliography}{nnn}\bibitem{Utiyama:1962sn}
\article[Utiyama:1962sn]{R. Utiyama, B.S. DeWitt}{J. Math. Phys.}{3}{608}{1962}
{Renormalization of a classical gravitational field interacting with quantized matter fields}.


\bibitem{Weinberg:1974tw}
\heparticle[Weinberg:1974tw]{S. Weinberg}{Problems in Gauge Field Theories}. 
\heparticle[Deser:1975nv]{S.~Deser}{The State of Quantum Gravity}.


\bibitem{Stelle:1976gc}
\article[Stelle:1976gc]{K.S. Stelle}{Phys. Rev.}{D16}{953}{1977}
{Renormalization of Higher Derivative Quantum Gravity}.


\bibitem{agravity}  \article[1403.4226]{A. Salvio, A. Strumia}{JHEP}{1406}{080}{2014}
{Agravity}.


\bibitem{WetterichInf} \article[1408.0156]{C. Wetterich}{Nucl. Phys.}{B897}{111}{2015}
{Inflation, quintessence, and the origin of mass}.

 


\bibitem{1502.01334}
\article[1502.01334]{K. Kannike, G. Hutsi, L. Pizza, A. Racioppi, M. Raidal, A. Salvio, A. Strumia}{JHEP}{1505}{065}{2015}
{Dynamically Induced Planck Scale and Inflation}.
\article[1509.05423]{K. Kannike, A. Racioppi, M. Raidal}{JHEP}{1601}{035}{2016}
{Linear inflation from quartic potential}.


\bibitem{Farzinnia:2015fka}
\article[1512.05890]{A. Farzinnia, S. Kouwn}{Phys. Rev.}{D93}{063528}{2016}
{Classically scale invariant inflation, supermassive WIMPs, and adimensional gravity}.


 


\bibitem{Salvio:2017xul}
\article[1703.08012]{A. Salvio}{Eur. Phys. J.}{C77}{267}{2017}
{Inflationary Perturbations in No-Scale Theories}.

\bibitem{Salles:2014rua}
\article[1401.4583]{F. d. O. Salles, I. L. Shapiro}{Phys. Rev.}{D89}{084054}{2014}{Do we have unitary and (super)renormalizable quantum gravity below the Planck scale?}.
   
\bibitem{Ivanov:2016hcm}
\article[1610.05330]{M. M. Ivanov, A. A. Tokareva}{JCAP}{1612}{018}{2016}
{Cosmology with a light ghost}.

\bibitem{Biswas:2005qr}
\article[hep-th/0508194]{T. Biswas, A. Mazumdar, W.~Siegel}{JCAP}{0603}{009}{2006}
{Bouncing universes in string-inspired gravity}.
 


\bibitem{Ostro}
M. Ostrogradski, {\em`` Memoire sur les equations differentielles relatives au probleme des isoperimetres''},
Mem. Ac. St. Petersbourg VI (1850) 385.


\bibitem{1512.01237}
\article[1512.01237]{A. Salvio, A. Strumia}{Eur. Phys. J.}{C76}{227}{2016}
{Quantum mechanics of 4-derivative theories}.
For later works see
\article[1611.03498]{M. Raidal, H. Veerm\"ae}{Nucl. Phys.}{B916}{607}{2017-03}
{On the Quantisation of Complex Higher Derivative Theories and Avoiding the Ostrogradsky Ghost}.
\article[1512.05305]{B. Holdom, J. Ren}{Phys. Rev.}{D93}{124030}{2016}
{QCD analogy for quantum gravity}.
\article[1608.01194]{A. Salvio}{Phys. Rev.}{D94}{096007}{2016}{Solving the Standard Model Problems in Softened Gravity}.
\heparticle[1703.04584]{D. Anselmi, M. Piva}{A new formulation of Lee-Wick quantum field theory}.
\heparticle[1703.05563]{D. Anselmi, M. Piva}{Perturbative unitarity of Lee-Wick quantum field theory}.
\heparticle[1704.01533]{J.F. Donoghue}{Quartic propagators, negative norms and the physical spectrum}.
\heparticle[1704.05031]{G. Narain}{Signs and Stability in Higher-Derivative Gravity}.
\heparticle[1704.07728]{D. Anselmi}{On the quantum field theory of the gravitational interactions}.


\bibitem{trini}
\article[1412.2769]{G.F. Giudice, G. Isidori, A. Salvio, A. Strumia}{JHEP}{1502}{137}{2015}
{Softened Gravity and the Extension of the Standard Model up to Infinite Energy}.
\article[1507.06848]{G.M. Pelaggi, A. Strumia, S. Vignali}{JHEP}{1508}{130}{2015}
{Totally asymptotically free trinification}.


\bibitem{1701.01453}
\heparticle[1701.01453]{G.M. Pelaggi, F. Sannino, A. Strumia, E. Vigiani}{Naturalness of asymptotically safe Higgs}.


\bibitem{Avramidi:1985ki}
I. G. Avramidi, A. O. Barvinsky, {\em``Asymptotic Freedom In Higher Derivative Quantum Gravity''},
Phys.\ Lett.\  B {159} (1985) 269 \doi{10.1016/0370-2693(85)90248-5}.

   

  
    


\bibitem{Avramidi:1986mj}
\heparticle[hep-th/9510140]{I.G. Avramidi}{Covariant methods for the calculation of the effective action in quantum field theory and investigation of higher derivative quantum gravity}.

\bibitem{deBerredoPeixoto:2004if}
\article[hep-th/0412249]{G. de Berredo-Peixoto, I. L. Shapiro}{Phys. Rev.}{D71}{064005}{2005}{Higher derivative quantum gravity with Gauss-Bonnet term}.
 
 
\bibitem{Julve:1978xn}
  J. Julve, M. Tonin,
   {\em``Quantum Gravity with Higher Derivative Terms''},
  Nuovo Cim.\ B 46 (1978) 137 \doi{doi:10.1007/BF02748637}.
  
\bibitem{Frad}
E. S. Fradkin, A. A. Tseytlin, {\em``Renormalizable asymptotically free quantum theory of gravity''},
Nucl. Phys. B 201 (1982) 469 \doi{10.1016/0550-3213(82)90444-8}.

\bibitem{Einhorn1}
\article[1511.01481]{M. B. Einhorn, D. R. T. Jones}{JHEP}{1601}{019}{2016}
{Induced Gravity I: Real Scalar Field}.
\article[1602.06290]{M. B. Einhorn, D. R. T. Jones}{JHEP}{1605}{185}{2016}
{Induced Gravity II: Grand Unification}.

 

  


\bibitem{Hooft:2010nc}
\heparticle[1011.0061]{G. 't Hooft}{The Conformal Constraint in Canonical Quantum Gravity}.


\bibitem{tHooft:2011aa}
\article[1104.4543]{G. 't Hooft}{Found.  Phys.}{41}{1829}{2011}
{A class of elementary particle models without any adjustable real parameters}.
  
  


\bibitem{Hooft:2015rdz}
\heparticle[1511.04427]{G. 't Hooft}{Singularities, horizons, firewalls, and local conformal symmetry}. 


\bibitem{Hamada:2002cm}
\article[Hamada:2002cm]{K. Hamada}{Prog. Theor. Phys.}{108}{399}{2002}
{Resummation and higher order renormalization in 4-D quantum gravity}.


\bibitem{Hamada:2009hb}
\article[0907.3969]{K. Hamada}{Found. Phys.}{39}{1356}{2009}
{Renormalizable 4D Quantum Gravity as A Perturbed Theory from CFT}.


\bibitem{Luo:2002ti}
For a recent summary see
\article[hep-ph/0211440]{M.X. Luo, H.W. Wang, Y. Xiao}{Phys. Rev.}{D67}{065019}{2003}{Two loop renormalization group equations in general gauge field theories}.


\bibitem{Mannheim:2009qi}
\article[0909.0212]{P.D. Mannheim}{Gen. Rel. Grav.}{43}{703}{2009}
{Comprehensive Solution to the Cosmological Constant, Zero-Point Energy, and Quantum Gravity Problems}.


\bibitem{Mannheim:2011ds}
\article[1101.2186]{P.D. Mannheim}{Found. Phys.}{42}{388}{2011}
{Making the Case for Conformal Gravity}.


\bibitem{ShapiroAnom}
E.~S.~Fradkin and A.~A.~Tseytlin,
 {\em ``Renormalizable asymptotically free quantum theory of gravity''},
  Nucl.\ Phys.\ B {201} (1982) 469
  \doi{10.1016/0550-3213(82)90444-8}.
  \article[hep-th/9205015]{I. Antoniadis, P.O. Mazur, E. Mottola}{Nucl. Phys.}{B388}{627}{1992}
{Conformal symmetry and central charges in four-dimensions}.
   I.~L.~Shapiro and A.~G.~Zheksenaev,
  {\em ``Gauge dependence in higher derivative quantum gravity and the conformal anomaly problem''},
  Phys.\ Lett.\ B {324} (1994) 286
  \doi{10.1016/0370-2693(94)90195-3}.
\article[hep-th/0307030]{G. de Berredo-Peixoto, I.L. Shapiro}{Phys. Rev.}{D70}{044024}{2003}
{Conformal quantum gravity with the Gauss-Bonnet term}.


\bibitem{Narain:2012te}
  \article[1211.5040]{G.Narain, R. Anishetty}{JHEP}{1307}{106}{2013}{Charge Renormalization due to Graviton Loops}.


\bibitem{Narain} \article[1309.0473]{G. Narain, R. Anishetty}{JHEP}{1310}{203}{2013}
{Running Couplings in Quantum Theory of Gravity Coupled with Gauge Fields}.
 
  
  


\bibitem{MV}
M.E. Machacek, M.T. Vaughn, {\em``Two Loop Renormalization Group Equations in a General Quantum Field Theory. 1. Wave Function Renormalization''},
Nucl. Phys. B 222 (1983) 83 \doi{10.1016/0550-3213(83)90610-7}.


\bibitem{Collins}
  L.~S.~Brown and J.~C.~Collins,
  {\em ``Dimensional Renormalization of Scalar Field Theory in Curved Space-time''},
  Annals Phys.\  {130} (1980) 215
  \doi{10.1016/0003-4916(80)90232-8}.
   S.~J.~Hathrell,
  {\em ``Trace Anomalies and $\lambda \phi^4$ Theory in Curved Space''},
  Annals Phys.\  {139} (1982) 136
  \doi{10.1016/0003-4916(82)90008-2}.
  
 


\bibitem{Hathrell}
 S.~J.~Hathrell,
  {\em ``Trace Anomalies and {QED} in Curved Space''},
  Annals Phys.\  {142} (1982) 34.
  \doi{10.1016/0003-4916(82)90227-5}. 
  M.~D.~Freeman,
  {\em``The Renormalization of Nonabelian Gauge Theories in Curved Space-time''},
  Annals Phys.\  {153} (1984) 339.
  \doi{doi:10.1016/0003-4916(84)90022-8}.


\bibitem{Osborn}
 I.~Jack and H.~Osborn,
  {\em ``Analogs for the $c$ Theorem for Four-dimensional Renormalizable Field Theories''},
  Nucl.\ Phys.\ B {343} (1990) 647
  \doi{10.1016/0550-3213(90)90584-Z}.
  


\bibitem{Tarasov:1980au}
  O.~V.~Tarasov, A.~A.~Vladimirov and A.~Y.~Zharkov,
  {\em ``The Gell-Mann-Low Function of QCD in the Three Loop Approximation''},
  Phys.\ Lett.\ B {93} (1980) 429 
  \doi{doi:10.1016/0370-2693(80)90358-5}.


\bibitem{Frohlich:1982tw}
\article[Frohlich:1982tw]{J. Frohlich}{Nucl. Phys.}{B200}{281}{1982}
{On the Triviality of $\lambda \phi^4$ in $D$ Dimensions Theories and the Approach to the Critical Point in $D\ge 4$ Dimensions}.
\article[Luscher:1987ay]{M. Luscher, P. Weisz}{Nucl. Phys.}{B290}{25}{1987}
{Scaling Laws and Triviality Bounds in the Lattice $\phi^4$ Theory. 1. One Component Model in the Symmetric Phase}.
\article[0902.3100]{U. Wolff}{Phys. Rev.}{D79}{105002}{2009}
{Precision check on triviality of $\phi^4$ theory by a new simulation method}.


\bibitem{wetterich}
C.~Wetterich, {\em ``Exact evolution equation for the effective potential''},  Phys.\ Lett.\ B {301} (1993) 90
  \doi{10.1016/0370-2693(93)90726-X}.
\article[hep-th/9403164]{T. Papenbrock, C. Wetterich}{Z. Phys.}{C65}{519}{1994}
{Two loop results from one-loop computations and nonperturbative solutions of exact evolution equations}.
\article[hep-ph/0005122]{J. Berges, N. Tetradis, C. Wetterich}{Phys. Rept.}{363}{223}{2000}
{Nonperturbative renormalization flow in quantum field theory and statistical physics}.


\bibitem{0811.2197}
\article[0811.2197]{A. Nicolis, R. Rattazzi, E. Trincherini}{Phys. Rev.}{D79}{064036}{2008}
{The Galileon as a local modification of gravity}.


\bibitem{Antoniadis}
I.~Antoniadis and E.~Mottola,
 {\em ``4-D quantum gravity in the conformal sector''},
  Phys.\ Rev.\ D {45} (1992) 2013
  \doi{10.1103/PhysRevD.45.2013}.
  \article[gr-qc/0612068]{I. Antoniadis, P.O. Mazur, E. Mottola}{New J. Phys.}{9}{11}{2006}
{Cosmological dark energy: Prospects for a dynamical theory}.
  \article[1606.08784]{M. Maggiore}{Fundam. Theor. Phys.}{187}{221}{2017}
{Nonlocal Infrared Modifications of Gravity. A Review}.

  


\bibitem{Sakharov}
{Sakharov's induced gravity: A Modern perspective}.
  K.~Akama, Y.~Chikashige, T.~Matsuki and H.~Terazawa,
   {\em ``Gravity and Electromagnetism as Collective Phenomena: A Derivation of Einstein's General Relativity''},
  Prog.\ Theor.\ Phys.\  {\bf 60} (1978) 868.
   S.~L.~Adler,
  {\em ``A Formula for the Induced Gravitational Constant''},
  Phys.\ Lett.\ B {95} (1980) 241.
   A.~Zee,
  {\em ``Spontaneously Generated Gravity''},
  Phys.\ Rev.\ D {23} (1981) 858.
 N.~N.~Khuri,
  {\em ``An Upper Bound for Induced Gravitation''},
  Phys.\ Rev.\ Lett.\  {49} (1982) 513.
  N.~N.~Khuri,
  {\em ``The Sign of the Induced Gravitational Constant''},
  Phys.\ Rev.\ D {26} (1982) 2664.
  S.~L.~Adler,
  {\em ``Einstein Gravity as a Symmetry Breaking Effect in Quantum Field Theory''},
  Rev.\ Mod.\ Phys.\  {54} (1982) 729
   [Rev.\ Mod.\ Phys.\  {55} (1983) 837].


\bibitem{Nima} See e.g.\ \article[hep-th/0210184]{N. Arkani-Hamed, H. Georgi, M.D. Schwartz}{Annals Phys.}{305}{96}{2002}
{Effective field theory for massive gravitons and gravity in theory space}.
 


\bibitem{SUSY}
G.~Veneziano and S.~Yankielowicz,
  {\em ``An Effective Lagrangian for the Pure N=1 Supersymmetric Yang-Mills Theory''},
  Phys.\ Lett.\  {113B} (1982) 231 \doi{10.1016/0370-2693(82)90828-0}.
\article[hep-th/9905015]{N.M. Davies, T.J. Hollowood, V.V. Khoze, M.P. Mattis}{Nucl. Phys.}{B559}{123}{1999}
{Gluino condensate and magnetic monopoles in supersymmetric gluodynamics}.
\article[hep-th/0309252]{F. Sannino, M. Shifman}{Phys. Rev.}{D69}{125004}{2003}
{Effective Lagrangians for orientifold theories}.


\bibitem{Shapiro:2000dz}
\article[hep-th/0012227]{I.L. Shapiro, J. Sola}{JHEP}{0202}{006}{2000}
{Scaling behavior of the cosmological constant: Interface between quantum field theory and cosmology}.


\end{thebibliography}
\end{document}